\documentclass[a4paper,fleqn,usenatbib]{mnras}

\usepackage{mathptmx}
\usepackage{afterpage}

\usepackage{graphicx}	
\usepackage{amsmath}	
\usepackage{amssymb}	

\newcommand{\VCYG}{V404 Cyg}
\newcommand{\SWIFT}{\textit{Swift}}
\newcommand{\MJD}[1]{MJD~#1}
\newcommand{\mita}[1]{{#1}}
\newcommand{\mrm}[1]{\mathrm{#1}}

\title[Polarization of V404 Cyg]{High-precision optical polarimetry of the accreting black hole\\ V404 Cyg during the June 2015 outburst}

\author[I. Kosenkov et al.]
{Ilia A. Kosenkov,$^{1,2,3}$\thanks{E-mail: ilia.kosenkov@utu.fi}
	Andrei V. Berdyugin,$^{1}$
	Vilppu Piirola,$^{1,4,5}$
    \newauthor Sergey S. Tsygankov,$^{1}$
       Enric Pall\'{e},$^{6}$ 
       Paulo A. Miles-P\'{a}ez$^{7}$
	  and  Juri Poutanen$^{1,2,8}$
	\\
	$^{1}$Tuorla Observatory, Department of Physics and Astronomy, University of Turku, V\"{a}is\"{a}l\"{a}ntie 20, FI-21500 Piikki\"{o}, Finland\\
	$^{2}$Nordita, KTH Royal Institute of Technology and Stockholm University, Roslagstullsbacken 23, SE-10691 Stockholm, Sweden\\
    $^{3}$Department of Astrophysics, St. Petersburg State University, Universitetskiy pr. 28, Peterhof, 198504 St. Petersburg, Russia\\  	$^{4}$Finnish Centre for Astronomy with ESO, University of Turku, V\"{a}is\"{a}l\"{a}ntie 20, FI-21500 Piikki\"{o}, Finland\\
  $^{5}$Kiepenheuer-Institut f\"{u}r Sonnenphysik, D-79104 Freiburg, Germany\\  
  $^{6}$Instituto de Astrofisica de Canarias, E-38205 La Laguna, Spain\\  
  $^{7}$The University of Western Ontario, Department of Physics and Astronomy, 1151 Richmond Avenue, London, ON N6A 3K7, Canada\\ 
  $^{8}$Kavli Institute for Theoretical Physics, University of California, Santa Barbara, CA 93106, USA 
}

\date{Accepted 2017 March 24. Received 2017 March 23; in original form 2017 February 07}
\pubyear{2017}

\begin{document}
	\label{firstpage}
	\pagerange{\pageref{firstpage}--\pageref{lastpage}}
	\maketitle
	
\begin{abstract}
Our simultaneous three-colour ($BVR$) polarimetric observations of the low-mass black hole X-ray binary \VCYG\ show a small but statistically significant change of polarization degree ($\Delta p \sim 1$ per cent) between the outburst in June 2015 and the quiescence. 
The polarization of \VCYG\ in the quiescent state agrees within the errors with that of the visually close (1\farcs4) companion 
($p_\mita{R} = 7.3\pm 0.1$ per cent), indicating that it is predominantly of interstellar origin. 
The polarization pattern of the surrounding field stars supports this conclusion. 
From the observed variable polarization during the outburst we show that polarization degree of the intrinsic component peaks in the $V$-band, 
$p_\mita{V} = 1.1\pm 0.1$ per cent, 
at the polarization position angle of 
$\theta_\mita{V}= -7\degr\pm 2\degr$,
which is consistent in all three passbands. 
We detect significant variations in the position angle of the intrinsic polarization in $R$ band from $-30\degr$ to $\sim 0\degr$ during the outburst peak. 
The observed wavelength dependence of the intrinsic polarization does not support non-thermal synchrotron emission from a jet as a plausible mechanism, but is in better agreement with the combined effect of electron (Thomson) scattering and absorption in a flattened plasma envelope or outflow surrounding the illuminating source. 
Alternatively, the polarization signal can be produced by scattering of the disc radiation in a mildly relativistic polar outflow. 
The position angle of the intrinsic polarization, nearly parallel to the jet direction (i.e. perpendicular to the accretion disc plane), is in agreement with these interpretations. 
	\end{abstract}
	
	


\section{Introduction}
\label{sec:intr}
		
\VCYG\ is a low-mass X-ray binary (LMXBs)  consisting of a black hole (BH) and a late K-type companion star with masses $(8-12){\rm M}_{\sun}$ and $(0.5-0.8){\rm M}_{\sun}$, respectively \citep{Shahbaz1994,Khargharia2010,Casares14}, with the error being dominated by the uncertainty in the orbital inclination. 
The distance to this source is known to a high precision $2.39\pm0.14\ \mrm{kpc}$ \citep{Miller-Jones2009}. 
\VCYG\ had been known as Nova Cyg 1938 and   went into outbursts also in 1956 and 1989 \citep{Richter89}. 
On the latter  occasion the source was also detected in the X-rays as an X-ray nova GS~2023+338 \citep{Makino1989}. 

One of the last outbursts of the source took place in 2015 June  \citep{Barthelmy2015} after 27 years of quiescence. 
In the peak of the outburst, the object showed very erratic behaviour with flares reaching a few tens of Crab in the hard X-ray domain;  accounting for  bolometric correction, this would correspond to the Eddington luminosity for a 10\,M$_{\sun}$ BH \citep{Segreto15,Rodriguez15a,Rodriguez15b}. 
The X-ray luminosity, however, still might be only a small fraction of the total one. 
The presence of the narrow iron K$\alpha$ line with the equivalent width of about 1 keV and a hard continuum imply that the central engine might not be observed directly but only through transmission in a powerful, optically thick outflows \citep{King15}. 
Thus the actual luminosity in the peak was likely in excess of Eddington. 
This interpretation is supported by the fast optical and X-ray variability \citep[][see Figs~\ref{fig:lc} and \ref{fig:xray}]{Kimura2016} that can be associated with the disc thermal instabilities as well as with the absorption in  the clumpy wind, both are characteristic of the accretion rates  approaching the critical Eddington value.

\begin{table*}
\centering
\caption{A log of observations of \VCYG. \label{tbl:obslog}
} 
\begin{minipage}{130mm}
\begin{tabular}{cccccccc}
	\hline
    	\hline
                   &  \multicolumn{2}{c}{$B$} &  \multicolumn{2}{c}{$V$} &  \multicolumn{2}{c}{$R$} & \\
	           MJD &              $p$  &       $\theta$  &            $p$    &        $\theta$ &               $p$ &       $\theta$  & Telescope\\
	               &        (per cent) &           (deg) &        (per cent) &           (deg) &        (per cent) &           (deg) & \\
	\hline
	      57195.16 &                -- &              -- &   $7.22 \pm 0.18$ &   $7.7 \pm 0.7$ &   $7.29 \pm 0.07$ &   $5.3 \pm 0.3$ & KVA\\ 
	      57196.15 &                -- &              -- &                -- &              -- &   $7.55 \pm 0.23$ &   $5.1 \pm 0.9$ & KVA\\ 
	      57197.15 &   $8.40 \pm 0.21$ &   $6.6 \pm 0.7$ &   $7.64 \pm 0.08$ &   $8.6 \pm 0.3$ &   $7.66 \pm 0.03$ &   $6.9 \pm 0.1$ & KVA\\ 
    	  57199.14 &   $9.57 \pm 0.55$ &   $8.0 \pm 1.6$ &   $7.28 \pm 0.10$ &   $9.1 \pm 0.4$ &   $7.31 \pm 0.06$ &   $7.5 \pm 0.2$ & KVA\\ 
    	  57200.14 &                -- &              -- &                -- &              -- &   $7.37 \pm 0.10$ &   $7.5 \pm 0.4$ & KVA\\ 
	      57206.01 &   $7.66 \pm 0.29$ &   $8.6 \pm 1.1$ &$6.63\pm0.09$&$10.5\pm0.4$&$7.13\pm0.05$&$7.9\pm0.2$& WHT\\ 
      	  57207.12 &   $8.28 \pm 0.43$ &   $6.7 \pm 1.5$ &$6.52\pm0.09$&$11.5\pm0.4$&$7.16\pm0.07$&$7.5\pm0.3$& WHT\\ 
    	  57208.11 &   $7.88 \pm 0.22$ &   $6.9 \pm 0.8$ &$6.68\pm0.11$&$11.0\pm0.5$&$7.16\pm0.08$&$7.4\pm0.3$& WHT\\ 
    	  57209.10 &   $7.94 \pm 0.49$ &  $11.8 \pm 1.8$ &$6.64\pm0.14$&$11.8\pm0.6$&$6.95\pm0.09$&$7.3\pm0.4$& WHT\\ 
	      57210.11 &   $7.38 \pm 0.75$ &   $8.4 \pm 2.9$ &$6.41\pm0.14$&$11.3\pm0.6$&$7.17\pm0.10$&$8.2\pm0.4$& WHT\\ 
          57651.37 &                -- &              -- &   $6.81 \pm 0.46$ &  $10.5 \pm 1.9$ &   $7.25 \pm 0.31$ &   $5.4 \pm 1.2$ & UH88\\
          57652.31 &                -- &              -- &   $8.23 \pm 0.60$ &  $8.5 \pm  2.1$ &   $7.49 \pm 0.26$ &   $8.5 \pm 1.0$ & UH88\\
   \multicolumn{8}{c}{Combined data} \\
   
57195--57200 &$8.55\pm0.20$&$6.7\pm0.7$&$7.47\pm0.06$&$8.6\pm0.2$&$7.51\pm0.03$&$6.8\pm0.1$& KVA\\
      57206--57210 &   $7.84 \pm 0.16$ &   $7.9 \pm 0.6$ &   $6.58 \pm 0.05$ &  $11.1 \pm 0.2$ &   $7.13 \pm 0.03$ &   $7.7 \pm 0.1$ & WHT\\
      57651--57652 &                -- &              -- &   $7.32 \pm 0.38$ &   $9.7 \pm 1.5$ &   $7.37 \pm 0.21$ &   $7.2 \pm 0.8$ & UH88\\
     \multicolumn{8}{c}{{Companion star}} \\ 
57206--57210   & -- & -- & $6.64\pm0.21$ & $11.9\pm0.9$ & $7.28\pm0.09$ & $8.5\pm 0.4$ & WHT\\
57651--57652          &  -- & -- & -- & -- & $7.25 \pm 0.35$ & $8.3 \pm 1.4$ & UH88\\
	\hline
\end{tabular}
\begin{flushleft}{ 
Note: The degree of polarization ($p$) and the position angle ($\theta$) in the $BVR$ bands are given for nightly averages as well as 
for the combined data, computed by weighted averaging of the Stokes parameters of individual observations. 
Errors are $1\sigma$.
}\end{flushleft} 
\end{minipage}
\end{table*}

Polarimetry is a powerful tool that can probe the geometry and the physical mechanisms producing optical -- near infrared (ONIR) radiation in LMXBs.
There were several reports of ONIR polarization observations of \VCYG\ during its outburst and shortly after it had ended \citep{Tanaka2016,Shahbaz2016,Itoh16}. 
However, the presented results do not agree with each other: \citet{Tanaka2016} argued in favour of non-variable polarization  predominantly of interstellar origin, while \citet{Shahbaz2016} observed  high and strongly variable intrinsic polarization. 
The main source of the optical emission of \VCYG\ in the  quiescent state is the late K-type star \citep{Hynes09,Khargharia2010} and therefore is unlikely to be polarized.  
Thus polarization observed in this state can be used to estimate the effects of the interstellar medium (ISM).  

During the outburst, the ONIR luminosity of the compact object increases by several orders of magnitude, completely outshining the second component of the system. 
The polarized emission can be produced by a number of mechanisms \citep[see also][]{Veledina13}.
First, radiation from the outer parts of the standard accretion disc, irradiated by the X-ray emission from the BH vicinity, can be polarized by scattering in  the disc atmosphere, with the polarization degree being dependent on the ratio of the absorption to scattering opacity \citep{Nagirner62}. 
Second, the jets detected in the BH X-ray binaries \citep[see review by][]{Fender14SSRv} including \VCYG\ (J. Miller-Jones et al., in prep.)  can become a source of variable polarized light \citep[e.g.][]{Zdziarski14} due to the nature of synchrotron radiation in ordered magnetic field.   
Third, the non-thermal electrons accelerated within the hot inner accretion flow may emit synchrotron radiation, which in principle  can be polarized  depending on the magnetic field structure \citep{PV14}.
Fourth, the accretion disc radiation can be scattered in the mildly relativistic outflow emanating from the central BH \citep{Beloborodov98,BP99}. 
Finally, as the X-ray luminosity of \VCYG\ was super-Eddington during flares in the outburst peak, powerful outflows can be formed \citep{SS73,PLF07} blocking  radiation produced in the BH vicinity and reprocessing  its large fraction into softer radiation in an extended envelope.  
There are also observational evidence that rather strong outflows are present in \VCYG\ \citep{King15,MunozDarias16Nat}.
The radiation from the outflow may be also polarized if the spherical symmetry is broken \citep{DGS95}. 

The aim of the present paper  was to  study  polarized  ONIR emission  with a high-precision polarimeter which does not have instrumental polarization.  
The high polarimetric accuracy and exceptional care is required in the case of \VCYG\ because of a large interstellar polarization and a need to carefully subtract it from the full signal to unveil a weaker intrinsic polarization.   
In Section~\ref{sec:obsda}, we present our polarimetric   observations in $BVR$ filters of \VCYG\ during and after its outburst in 2015 June  complemented by polarimetry of a sample of the field stars. 
We also present the X-ray data obtained with {\it Swift} satellite. 
Section~\ref{sec:res} is devoted to the analysis of the polarimetric data of the source and of the field stars with the subsequent derivation of the intrinsic polarization of \VCYG.  
We discuss the obtained results and compare them with previously published polarimetric data in Section~\ref{sec:disc}, where we also present our interpretation of the results. 
We conclude in Section~\ref{sec:concl}.

\section{Observations and Data Analysis}
\label{sec:obsda}
		
\subsection{Polarimetry}
        
\begin{figure}			
	\includegraphics[keepaspectratio, width = 1\linewidth]{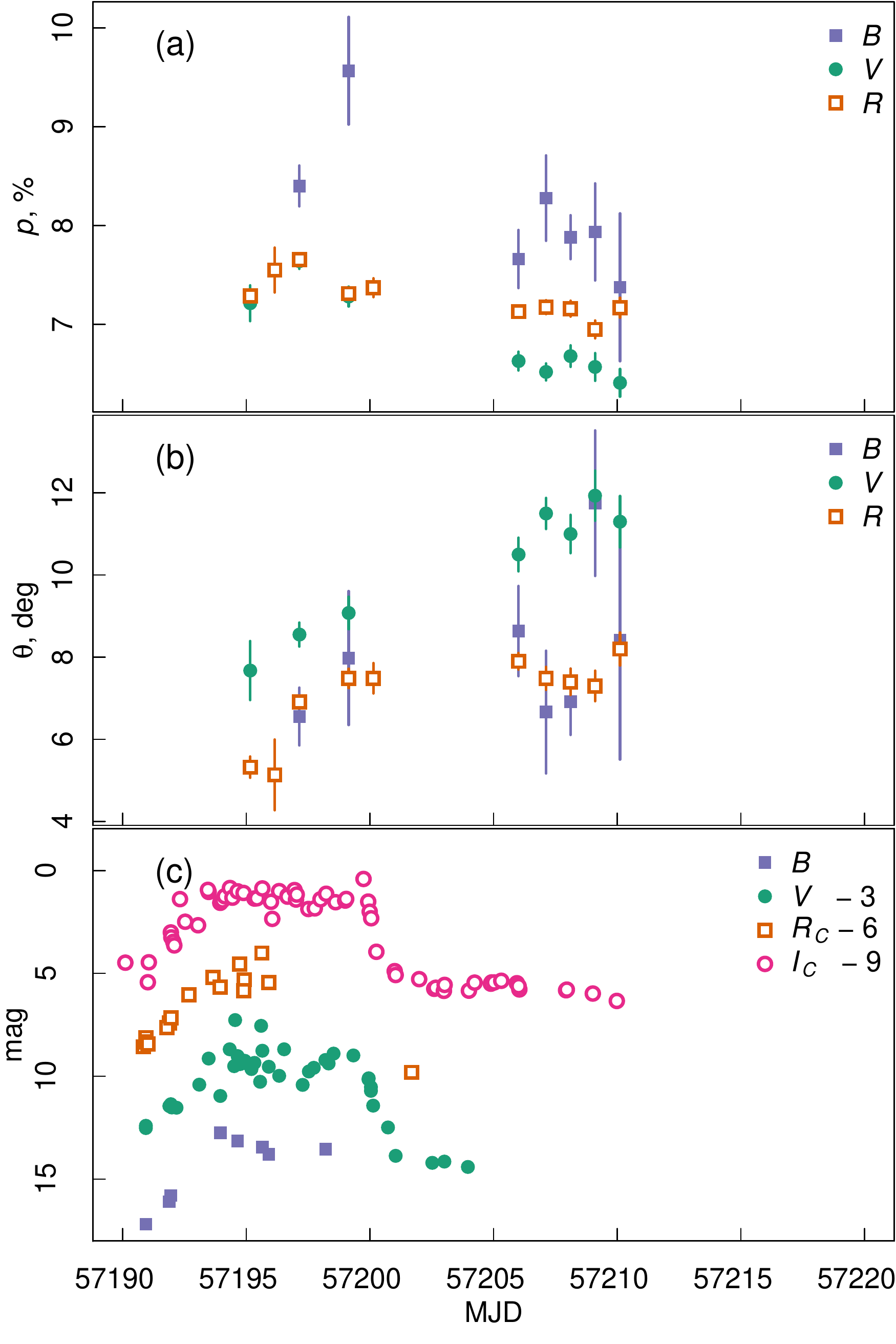}
	\caption{The 2015 June-July outburst of \VCYG. 
From top to bottom: panels (a) and (b) show the observed polarization degree and polarization position angle of \VCYG\ in three bands; panel (c) gives the ONIR light curves from \citet{Kimura2016} (error bars are smaller than the symbols).
}
	\label{fig:lc}
\end{figure}

We have made polarimetric observations of \VCYG\ during the outburst on 5 nights in 2015 June 21--26 with the KVA 60~cm telescope at Observatorio del Roque de los Muchachos (ORM), La Palma, and during quiescence on five nights in 2015 July 2--7 with the 4.2~m William Herschel Telescope (WHT) at ORM using the polarimeter Dipol-2 \citep{piirola14}. 
Polarimetry of \VCYG\ was carried out in the quiescence also at the 2.2~m University of Hawai'i telescope (UH88) at Mauna Kea on two nights of 2016 September 19--20, using another copy of Dipol-2, identical to that used at ORM.
Journal of the observations of \VCYG\ is given in Table~\ref{tbl:obslog} and the main results are shown in Fig.~\ref{fig:lc}.  

The polarimeter, Dipol-2, is capable of making simultaneous measurements in three passbands ($BVR$) with high sensitivity. 
The detection limit of polarization is at the level of $10^{-5}$, set in practice by photon noise. 
An important asset of the instrument is that the sky background polarization is directly (optically) eliminated. 
The perpendicularly polarized components of sky are superimposed by the plane parallel calcite beam splitter, and sky polarization is thereby canceled \citep{Piirola1973}. 
This is essential, as the polarized flux from scattered skylight can exceed by orders of magnitude the signal from the target, particularly in bright Moon conditions. 
Dipol-2 has been found to be very stable and reliable instrument as demonstrated recently by detection of the variable polarization at 0.1 per cent level from a massive binary HD~48099 \citep{BPS16}.
 
The Dipol-2 polarimetry routine consists of cycles of 16 exposures at different orientations of the superachromatic half-wave retarder (22\fdg5 steps), corresponding to a full (360\degr) rotation of the retarder. 
Each successive 4 exposures give one independent measurement of the normalized Stokes parameters $q$ and $u$. 
Accordingly, one cycle provides four independent measurements of $q$ and $u$. 
Typically, either two or four cycles were carried out on each observing night for \VCYG, yielding either 8 or 16 measurements of $q$ and $u$, used then to compute the nightly average polarization. 
The error estimate of the polarization degree was computed as the quadratic mean of the standard errors of the mean values for $q$ and $u$: 
\begin{equation}\label{eq:errorp}
\epsilon_{p} = \sqrt{(\epsilon_{q}^2 + \epsilon_{u}^2)/2} .
\end{equation} 
This formula is valid under conditions when the normalized Stokes parameters $q$ and $u$ are measured simultaneously and $\epsilon_\mita{p} \ll p$, which is always the case in our observations of \VCYG. 
In the ideal case, when $\epsilon_{q} = \epsilon_{u}$ \citep[see e.g.][]{serkowski62,clarke71} equation (\ref{eq:errorp}) gives $\epsilon_{p} = \epsilon_{q} = \epsilon_{u}$. 
The error estimate for the polarization angle can then be  expressed in terms of polarization degree and its error as 
\begin{equation}\label{eq:errortheta}
\epsilon_\theta =  28\fdg 65\   \epsilon_{\mita{p}} / {p}.
\end{equation}  
With typical exposure times of 15~s on the WHT and UH88 and 30--45~s on the KVA and a $\sim$3~s readout time between the exposures (3 cameras operating simultaneously), the total telescope time used for \VCYG\ on each night was about 10--20 min at WHT and UH88 and 20--40 min at KVA.

Standard CCD reduction procedures (bias and dark subtraction, flat fielding) were applied prior to extracting the fluxes from the double images of the target, formed onto the CCD by the polarizing calcite beam splitter. 
Special centering algorithm and subframing procedures were used to facilitate processing a large number, up to several hundred, of exposures at the same time. 
In computing the mean values of $q$ and $u$ we applied a ``2$\sigma$'' iterative weighting algorithm. The initial mean and standard deviation were obtained applying equal weights to all points. Then on each step individual points deviating more than two standard deviations from the mean ($d > 2 \sigma$) were given a lower weight, proportional to the inverse square of the error estimate, $\epsilon_{x}$. 
The value $\epsilon_{x} = \sigma$ for $d < 2 \sigma$ was assumed to increase linearly from $\epsilon_{x} = 1\sigma$ to $3 \sigma$ with $d$ increasing from $2\sigma$ to $3 \sigma$. 
Points with $d > 3 \sigma$ were rejected. The procedure converges fast and values of mean and standard deviation are obtained within a few iterations.
Under normal conditions, 6--8 per cent of individual points deviated more than $2\sigma$ and were given lower weight ($W<1$). 
The remaining 92--94 per cent of points were equally weighted ($W=1$). 
The weighting procedure helps to suppress effects from transient clouds, moments of bad seeing, cosmic ray events, etc.

The instrumental polarization produced by the telescope was determined by observing a number (10--20) of unpolarized nearby stars. 
The Stokes parameters of the instrumental polarization ($q_\mrm{tel}, u_\mrm{tel}$) were obtained as the average from the sample, with an uncertainty of less than $3\times 10^{-6}$ in each passband, and were subtracted from the measured values of the Stokes parameters $q$ and $u$. 
For each of the three telescopes, Cassegrain focus was used and the instrumental polarization was negligible (less than $10^{-4}$) in the present context. 
Polarization position angle zero-point was determined by observations of large polarization standard stars: HD25443, HD161056, HD204827, BD+25\,727, and BD+59\,389. 
We also checked for possible scale correction factors of the polarization degree, and found evidence of small calibration coefficients (1.02--1.04) needed in the $V$ and $R$ passbands. 
Though these differences can be partially due to systematic errors in the published values  \citep{Hsu82,Turnshek90,Schmidt92}, we have applied the corrections to bring our data into the system commonly used by other investigators. 
Careful determination of the angle zero-point rules out a possibility of systematic difference in determination of polarization angle made with the different telescopes.

\begin{table}
\centering
\caption{Coordinates of the observed field stars in the direct vicinity of \VCYG.}
\label{tbl:fieldcoor}
\begin{tabular}{rccc}
	\hline
    	\hline
	Star & RA (J2000) & Dec. (J2000) & Name \\
	\hline
    	404 &   20:24:03.83 &   33:52:02.2 &                           \VCYG \\ 
	   4040 &   20:24:03.80 &   33:52:03.6 &                 Close companion \\ 
	   4041 &   20:24:03.00 &   33:51:29.3 &                                 \\ 
	   4042 &   20:23:57.13 &   33:52:39.3 &                                 \\ 
	   4043 &   20:24:10.23 &   33:53:02.8 &                                 \\ 
	   4044 &   20:23:49.22 &   33:50:09.4 &                                 \\ 
	   4045 &   20:23:49.69 &   33:48:24.6 &                                 \\ 
	   4046 &   20:24:12.84 &   33:50:11.7 &                                 \\ 
	   4047 &   20:24:19.96 &   33:47:44.1 &                                 \\ 
	   4048 &   20:23:49.92 &   33:54:49.7 &                                 \\ 
	   4049 &   20:24:19.63 &   33:52:42.4 &                                 \\ 
	  40405 &   20:24:02.35 &   33:54:04.8 &         USNO-B1.0 1239-00424177 \\ 
	  40408 &   20:23:43.00 &   33:51:12.9 &         USNO-B1.0 1238-00434864 \\ 
	  40409 &   20:24:25.29 &   33:53:24.4 &                 TYC 2693-1473-1 \\ 
	  40410 &   20:24:28.28 &   33:51:13.2 &                 TYC 2693-1457-1 \\ 
	  40411 &   20:23:59.44 &   33:46:53.9 &                     EM* VES 209 \\ 
	  40412 &   20:24:00.09 &   33:46:35.3 &         2MASS J20240008+3346353 \\ 
	  40413 &   20:24:20.44 &   33:56:25.7 &                        HD332228 \\ 
	  40416 &   20:24:24.17 &   33:47:17.7 &                 TYC 2693-1483-1 \\ 
	  40417 &   20:24:18.09 &   33:57:56.6 &                  TYC 2693-483-1 \\ 
	  40418 &   20:23:39.00 &   33:56:33.0 &                  TYC 2680-177-1 \\ 
	  40420 &   20:23:26.24 &   33:50:05.2 &                  TYC 2680-419-1 \\ 
	  40421 &   20:24:32.62 &   33:57:59.0 &                  TYC 2693-573-1 \\ 
	  40423 &   20:23:49.02 &   33:43:15.5 &                    TYC 2676-7-1 \\ 
	  40425 &   20:23:17.95 &   33:51:57.6 &                  TYC 2680-329-1 \\ 
	  40427 &   20:23:29.85 &   33:45:19.6 &                  TYC 2680-269-1 \\ 
	\hline
\end{tabular}
\end{table}

To study interstellar polarization in the direction of \VCYG, we have measured polarization of a sample of stars in the $10\arcmin\times 10\arcmin$ area of the sky around \VCYG\ (see Table~\ref{tbl:fieldcoor}). 
These observations have been done with the UH88 telescope in 2016 June 16-27 and September 18--25. 
Exposure time was set in the range of 5--15~s, depending on the brightness of the star. 
Mean values of Stokes parameters $q$ and $u$ for each star have been obtained by averaging 8--16 single measurements. 
Polarization of the close visual companion star, located at 1\farcs4 north of \VCYG\ \citep{Udalski91}, was measured on four nights of the WHT observing run (Table~\ref{tbl:obslog}). 
For very faint and red stars, polarization in the $B$-band cannot be determined with sufficient level of confidence.         
We emphasize that our study of polarization of \VCYG\ at the outburst and after it, as well as the observations of the field stars, have been done with the same instrument. 
The same methods of observation, calibration and data reduction have been applied to all data allowing us to avoid any systematic biases and/or offsets which may appear when polarization data obtained with the different instruments at different wavelengths are combined together.   
This is a major difference between our and recently published studies of polarization in \VCYG\ \citep{Tanaka2016,Shahbaz2016}.

\begin{figure}
\includegraphics[keepaspectratio, width = 1\linewidth]{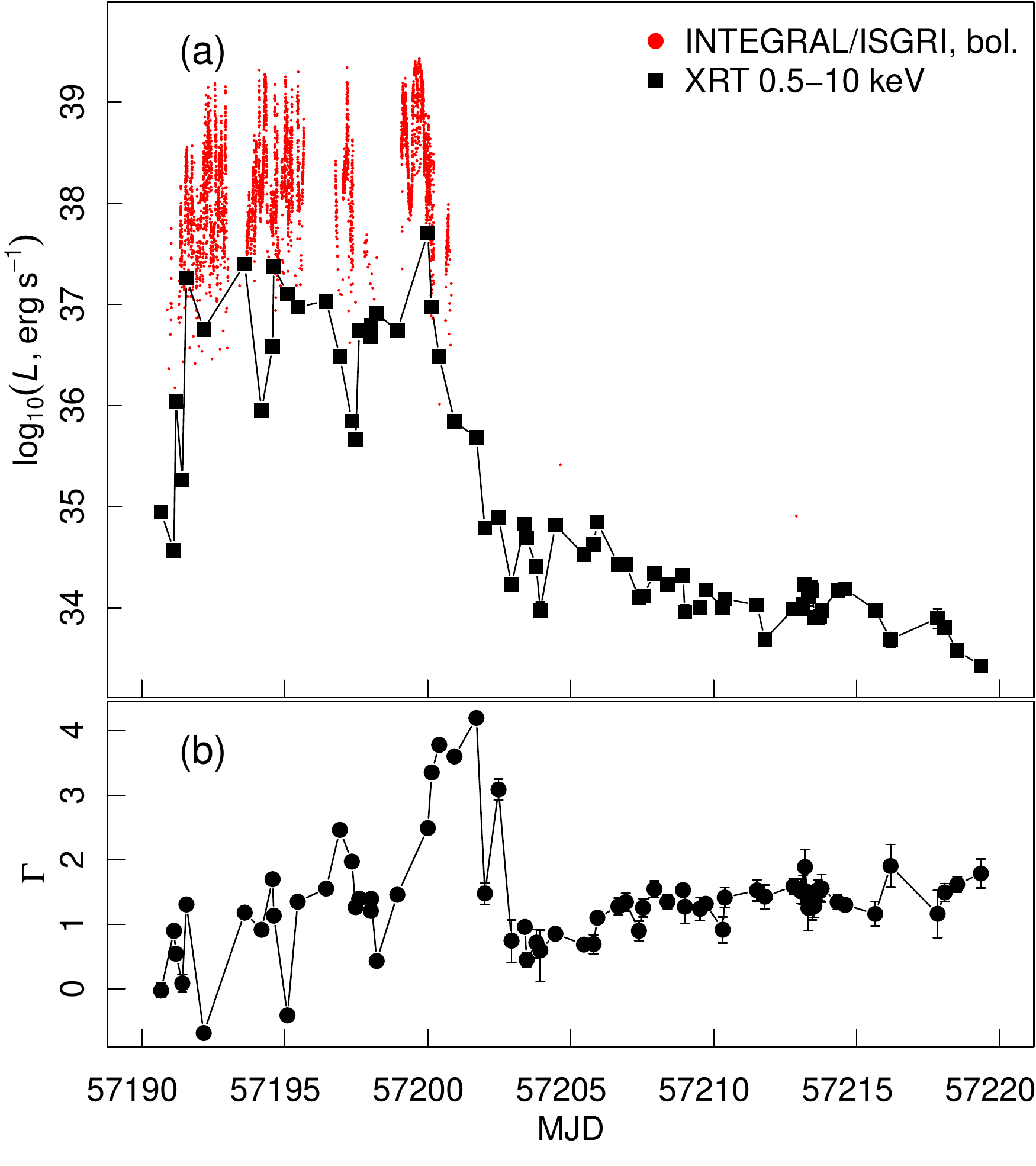}
\caption{The X-ray light curves of  \VCYG. 
Panel (a) shows unabsorbed X-ray luminosity measured with {\it Swift}/XRT in the 0.5--10 keV band (black squares) as well as the bolometric luminosity estimated from the {\it INTEGRAL} IBIS/ISGRI count rate in the 25--60 keV band (see text for the details).
Panel (b) shows the photon index $\Gamma$ as measured in the \SWIFT/XRT band.
}
\label{fig:xray}
\end{figure}

		
\subsection{X-ray data}
\label{subsec:xray}
			
\VCYG\ shows fast variability in all spectral ranges. 
To search for possible correlations with polarization, we were interested in monitoring data in both soft and hard X-ray energy bands. 
Such data were collected by the \SWIFT\ observatory \citep{2004ApJ...611.1005G} and the {\it INTEGRAL} \citep{2003A&A...411L...1W} which monitored \VCYG\ during the outburst.
			
The \SWIFT/XRT telescope \citep{2005SSRv..120..165B} covering soft X-ray band (0.2--10 keV) performed observations both in photon counting (PC) and windowed timing (WT) modes. 
We used all observations during the active phase of the outburst available in the archive. 
The spectrum for each observation was extracted using the online tools provided by the UK Swift Science Data Centre \citep{2009MNRAS.397.1177E}.\footnote{\url{http://www.swift.ac.uk/user_objects/}}
The obtained spectra were grouped to have at least one count per bin and were fitted in 0.5--10~keV band in {\sc XSPEC} package using Cash statistics \citep{1979ApJ...228..939C}.  
The simplest spectral model (absorbed power-law, i.e. {\sc phabs$\times$power} in {\sc xspec}) was used for the fitting procedure. 
The hydrogen column density $N_\mrm{H}$ was found to be variable between  $\sim0.5\times10^{22}$ and $\sim5\times10^{22}$ cm$^{-2}$. 
The latter value is higher than the Galactic absorption value in this direction, $N_\mrm{H}=0.7\times10^{22}$~cm$^{-2}$, given by \citet{2005A&A...440..775K}. 
However, an independent determination of the absorption value and photon index $\Gamma$ results in a substantial increase of uncertainty on both parameters due to narrowness of the \SWIFT/XRT energy band. 
On the other side, we checked that the X-ray flux (see Fig.~\ref{fig:xray}(a)) depends on  $N_\mrm{H}$ only weakly and the Galactic absorption value can be used for the robustness. 
Such approach is valid for this work because we are interested only in general trends in the X-ray flux and its correlation with the polarization behaviour. 
\VCYG\ shows strong variability of the X-ray spectrum (see Fig.~\ref{fig:xray}(b))  being first very hard, then becoming very soft by the end of the outburst around \MJD{57200--57202} and suddenly becoming hard again on \MJD{57203} with the simultaneous drop of the X-ray flux by orders of magnitude, which would correspond to the transition to the hard state \citep{ZG04}. 

In hard X-ray band (25--60~keV) \VCYG\ was monitored by the {\it INTEGRAL} IBIS/ISGRI telescope \citep{2003A&A...411L.131U} during the revolutions 1554--1563 (MJD 57190--57216). 
The light curves in the 25--60~keV energy range were obtained from the INTEGRAL Science Data Centre\footnote{\url{http://www.isdc.unige.ch/integral/Operations/Shift/QLAsources/V404_Cygni/V404_Cygni.php}; \citet{2015ATel.7758....1K}.} and converted to the bolometric luminosity  assuming a constant count rate from the Crab Nebula of 147~cnt\,s$^{-1}$ in the same energy range and a bolometric correction factor of $L_{\rm bol}/L_{25-60}=9.97$ \citep{Kimura2016}. 
We see strong variations in the hard X-ray flux in 25--60 keV energy band by almost three orders of magnitude on the timescale of hours (see red dots in Fig.~\ref{fig:xray}(a)), with the peak (bolometric) luminosities reaching the Eddington value for a 10\,M$_{\sun}$ BH.

\section{Results}
\label{sec:res}

\subsection{Polarization of \VCYG}
\label{subsec:pvcyg}
            
Fig. \ref{fig:lc} shows the degree of polarization, $p$, and the polarization position angle, $\theta$, of \VCYG\ in the $BVR$ bands, as well as the light curves in several passbands. 
The first set of our polarization observations was obtained with the 60 cm KVA telescope during the peak of the outburst (\MJD{57195--57200}). 
The brightness of the source was in the range of  $m_\mita{V} = 11.5 - 14.5$, strongly variable on the time scale of several hours  \citep{Kimura2016}. 
Nearly equal average polarization is observed in the $R$ and $V$ bands,
$p_\mita{R} = 7.47\pm 0.06$ 
 and $p_\mita{V} = 7.51 \pm 0.03$ per cent, 
while in the $B$ band $p$ is significantly larger (up to $9.57\pm 0.55$ per cent on \MJD{57199}).
The position angle $\theta$ has increased by $1\degr-2\degr$ during these 5 days. 

\begin{figure}
			\begin{minipage}{1\linewidth}
            	\includegraphics[keepaspectratio, width = 1\linewidth]{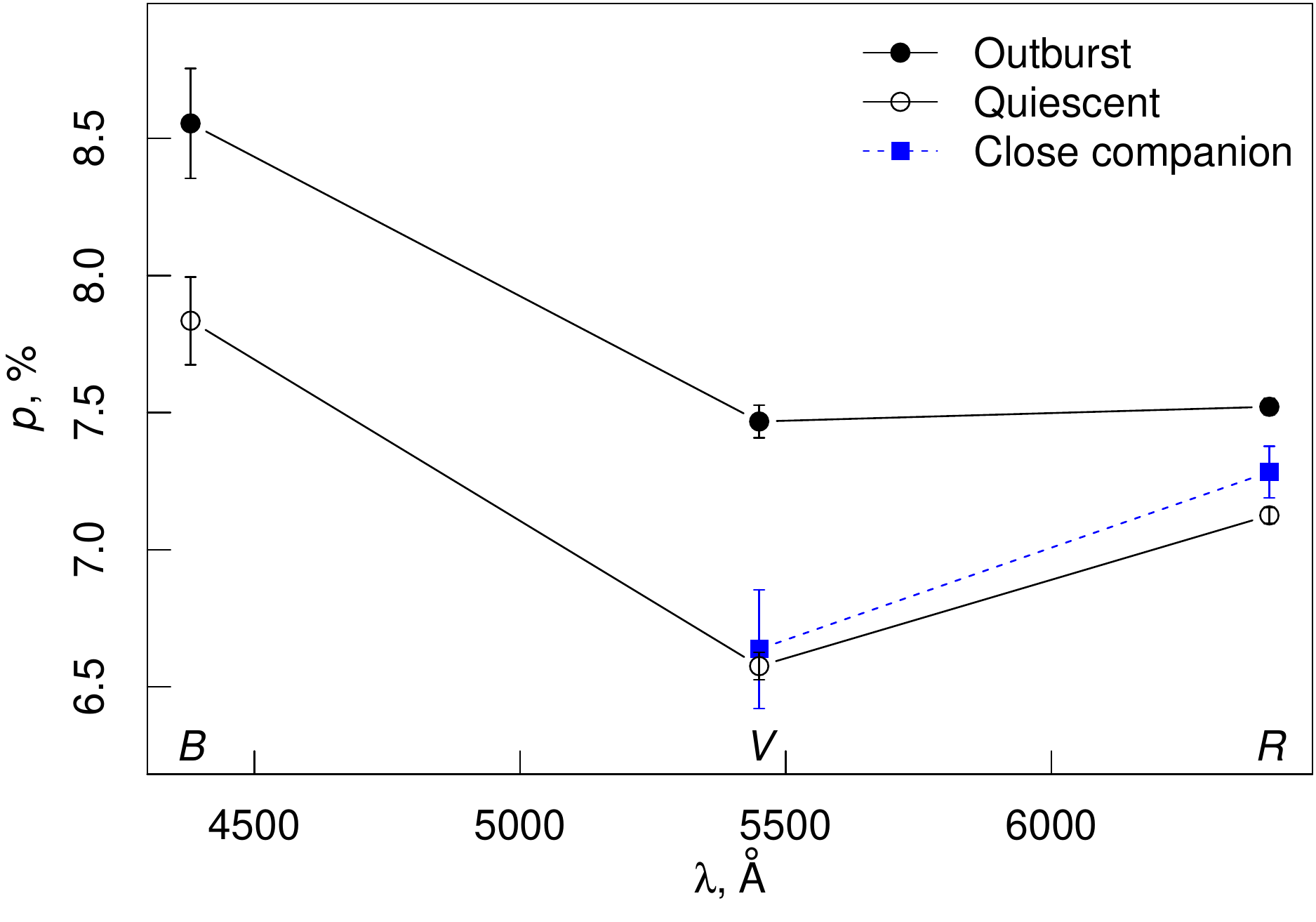}
			\end{minipage}
			\vfill
			\begin{minipage}{1\linewidth}
				\includegraphics[keepaspectratio, width = 1\linewidth]{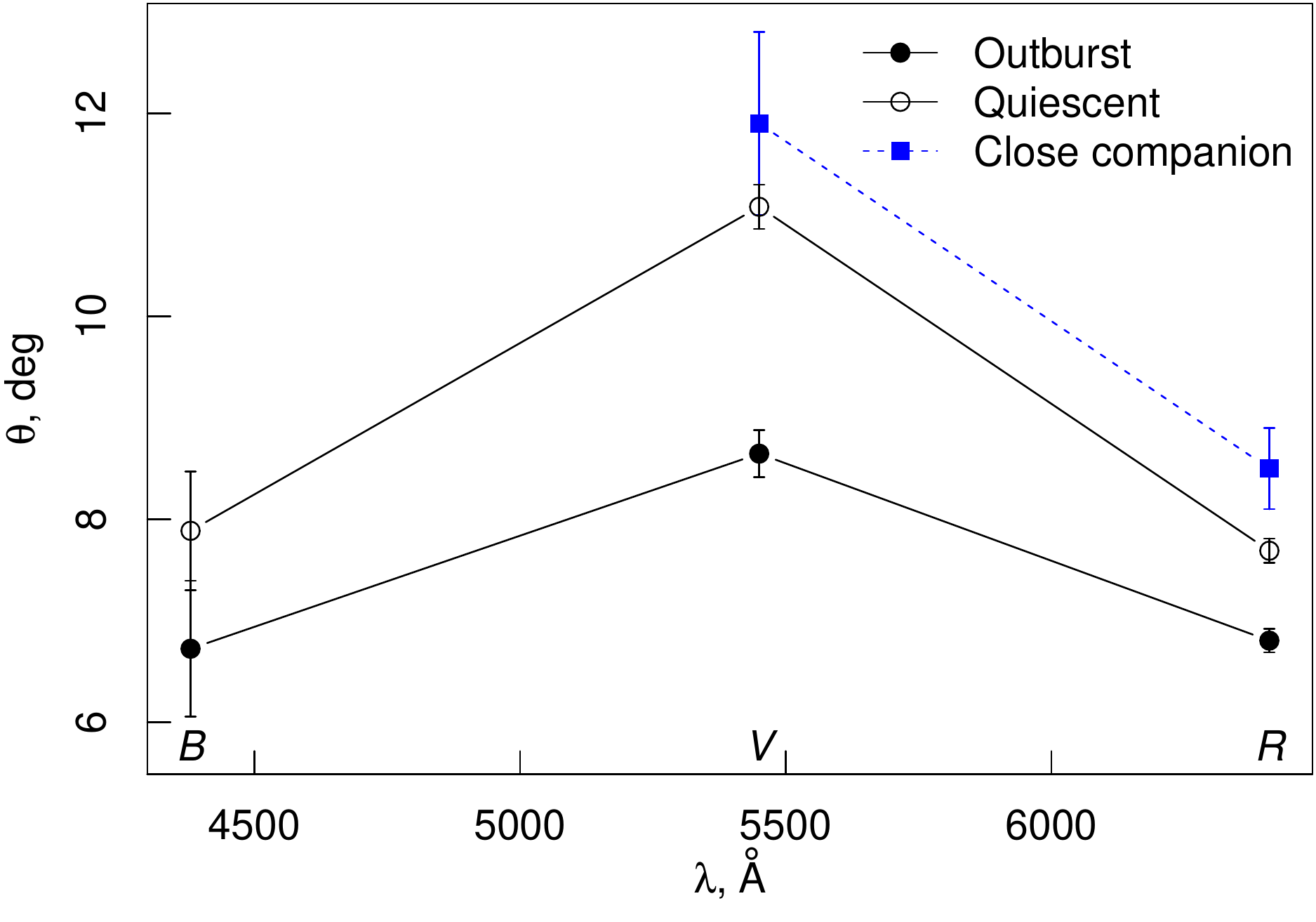}
			\end{minipage}
\caption{Average polarization degree (top) and position angle (bottom) of \VCYG\ (combined KVA data during the outburst and WHT data in quiescence) and its close companion.           
The values for \VCYG\  during the outburst and  in quiescence are shown by  filled and open black circles, respectively; the companion star is shown with the blue squares.  
}
			\label{fig:V404PL}
		\end{figure}

Polarimetric observations continued on \MJD{57206--57210} at the WHT when the source was already in the quiescent state: the ONIR brightness had dropped by 5 magnitudes and the X-ray flux was lower by 3 orders of magnitude  (Fig.~\ref{fig:lc}c and Fig.~\ref{fig:xray}a). 
The second set of observations gives evidence of a change between the outburst and the quiescence, both in the degree of polarization and the position angle, with $p_\mita{B}$, $p_\mita{V}$ and $p_\mita{R}$ reduced to 
$7.84\pm 0.16$, $6.58\pm 0.05$, $7.13\pm 0.03$ per cent on \MJD{57206--57210},
respectively. 
			
\begin{figure*}
					 	\begin{minipage}{0.495\linewidth}
					 		\includegraphics[width = 1\linewidth, keepaspectratio]{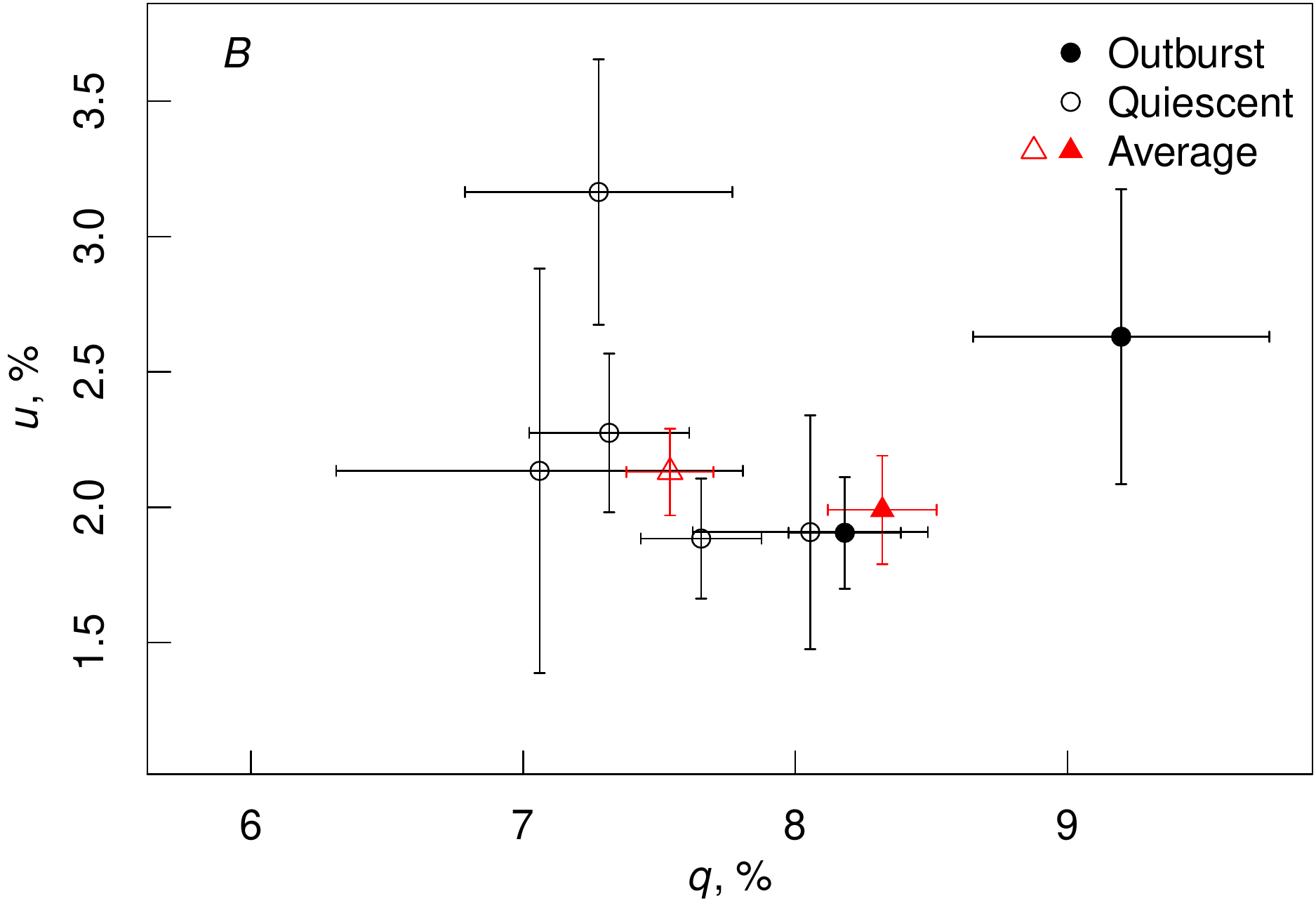}
					 	\end{minipage}
					 	\begin{minipage}{0.495\linewidth}
					 		\includegraphics[width = 1\linewidth, keepaspectratio]{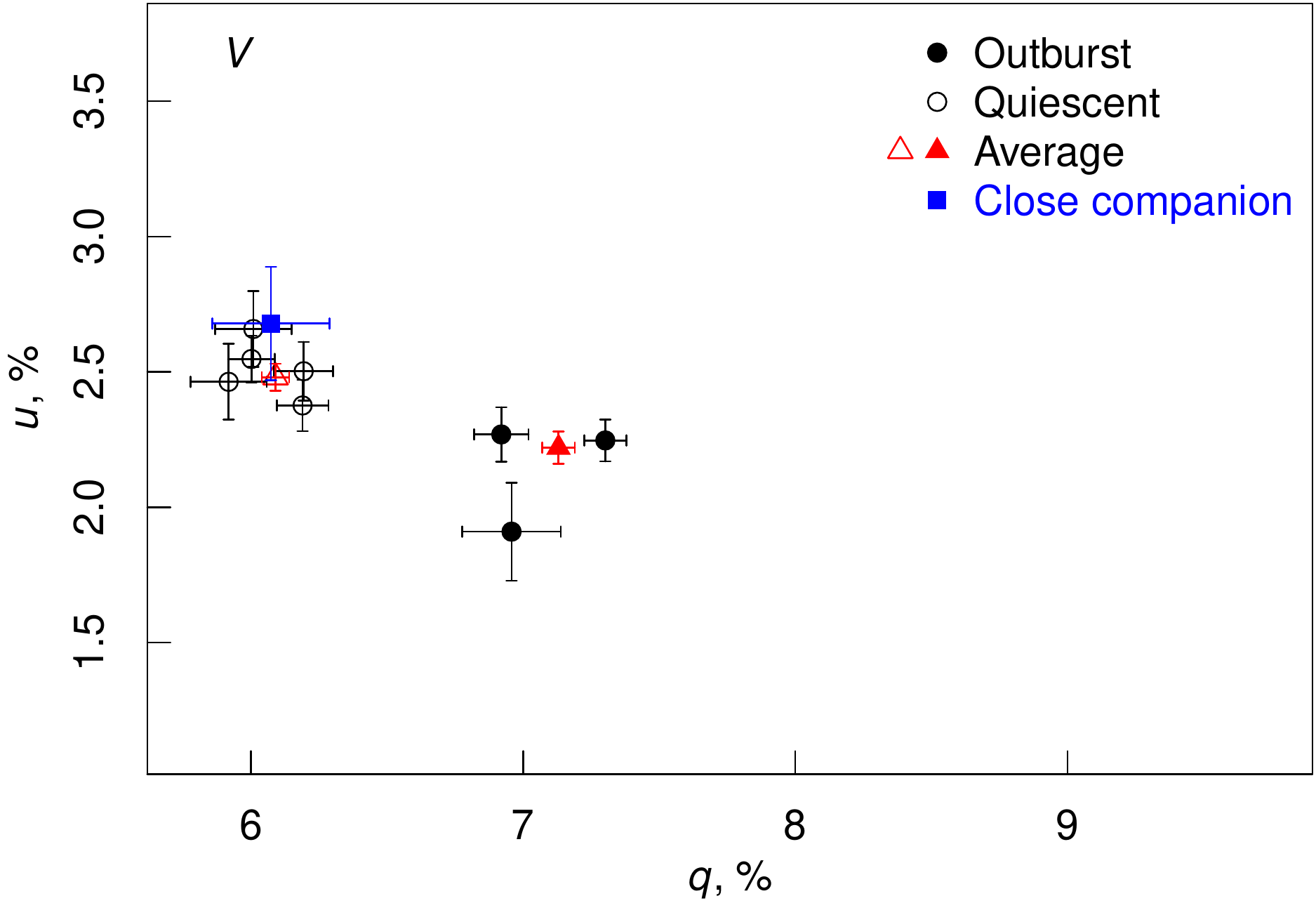}
					 	\end{minipage}
					 	\begin{minipage}{0.495\linewidth}
                            \includegraphics[width = 1\linewidth, keepaspectratio]{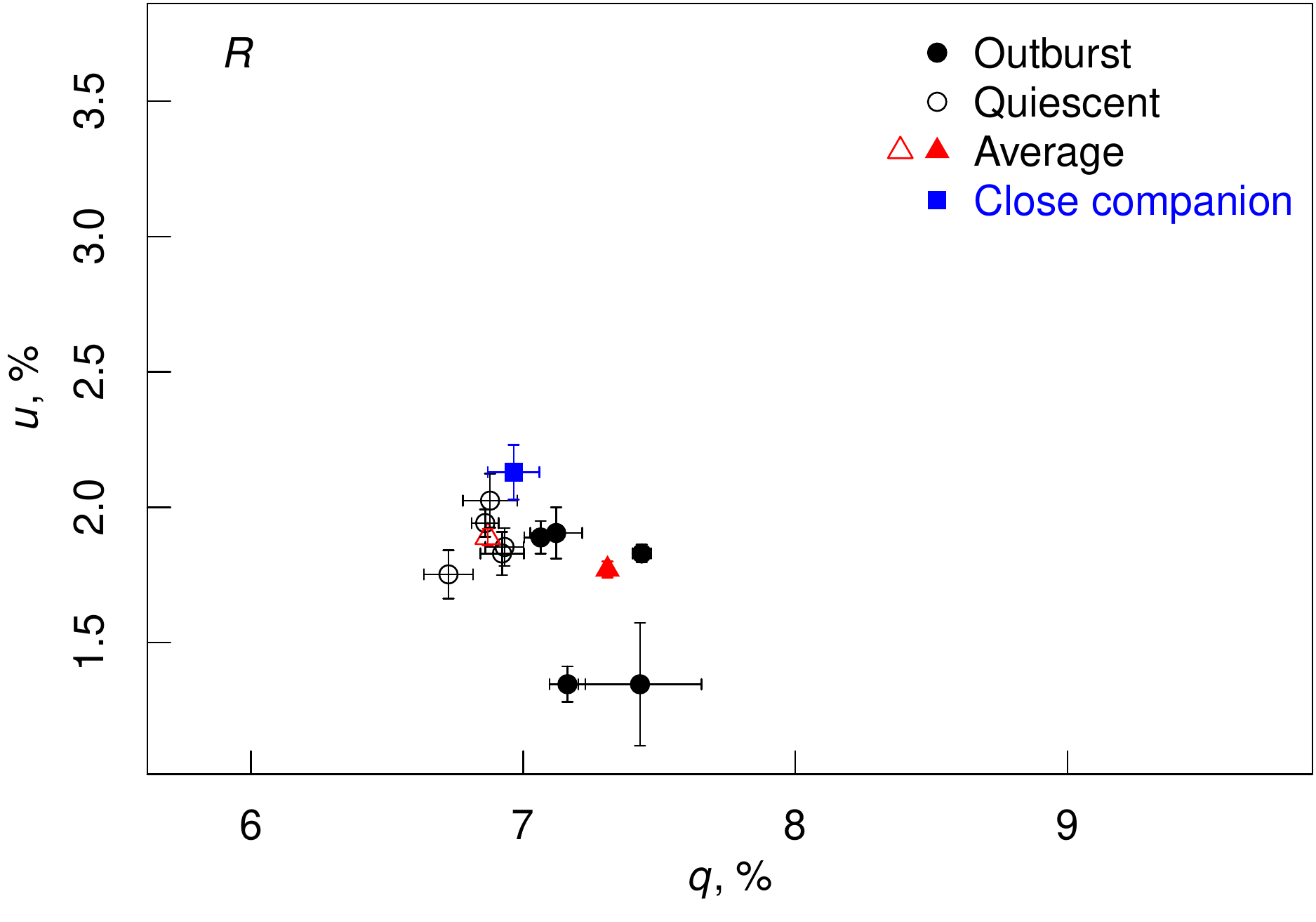}
					 	\end{minipage}
					 	\begin{minipage}{0.495\linewidth}
                            \includegraphics[width = 1\linewidth, keepaspectratio]{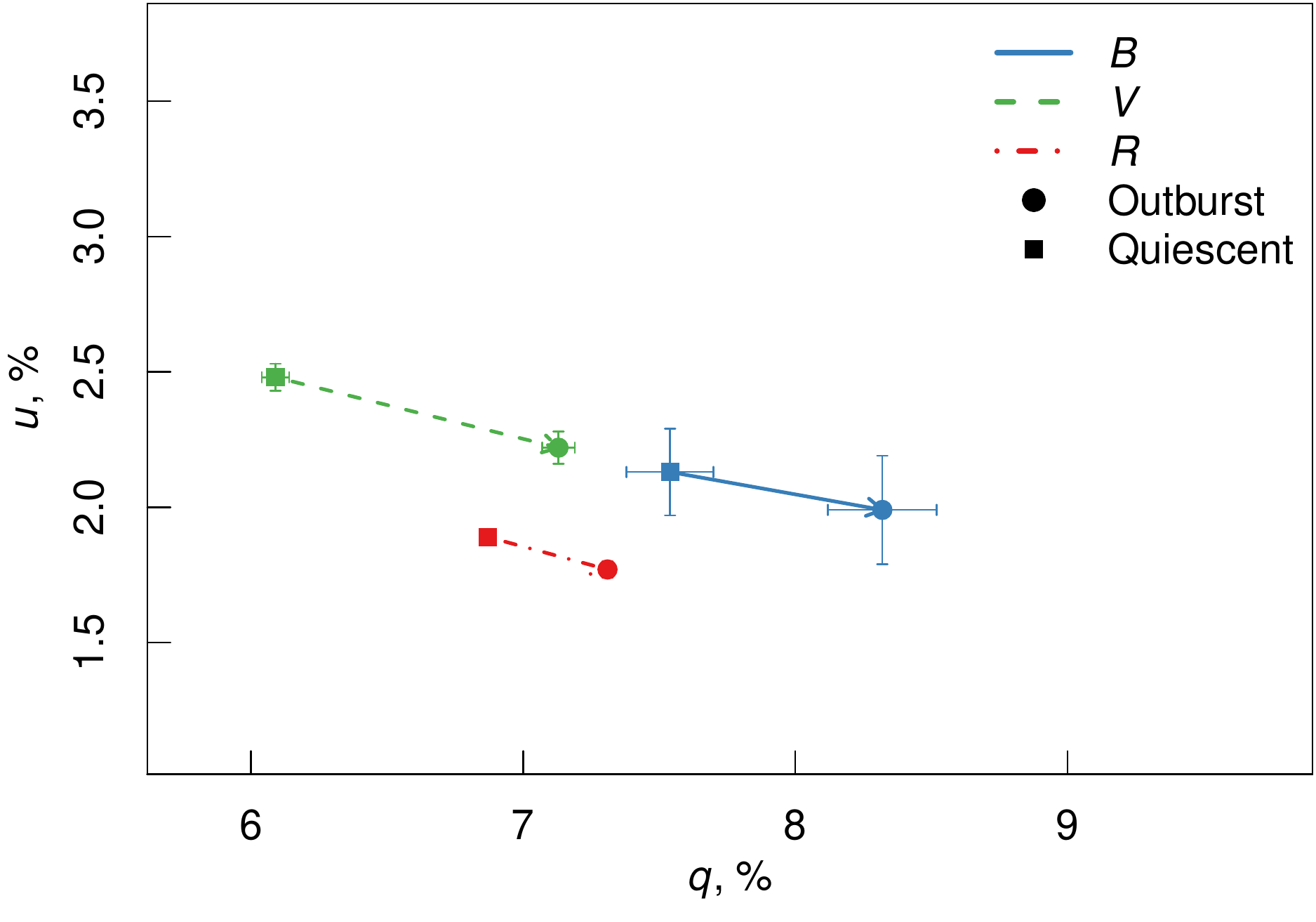}
					 	\end{minipage}
\caption{
($q,u$)-diagrams of \VCYG\ in the $B$, $V$, and $R$ bands  (from left to right and  from top to bottom). 
The filled black circles show the day-by-day data for the outburst, while the black open circles correspond to the quiescence. 
The red filled/open triangles give the average values for the outburst/quiescent states, respectively,  while the 
blue square gives the data for the close companion.
The lower right panel presents the average ($q,u$) values during the outburst (circles) and in quiescence (squares) with the arrows indicating the change in polarization of \VCYG.  
}
\label{fig:QU}
\end{figure*}

The wavelength dependence of $p$ and $\theta$ of \VCYG\ shows a peculiar profile with a significant dip in polarization longward of the $B$-band, as can be seen in Fig.~\ref{fig:V404PL}, where the average values of $p$ and $\theta$ during and after the outburst are presented. 
\VCYG\ polarization in quiescence is remarkably similar to that of its close companion, for which we have obtained measurements in the $V$ and $R$ bands (Table \ref{tbl:obslog}). 
This indicates that the polarization of \VCYG\ in quiescence is very likely of interstellar origin (Fig. \ref{fig:V404PL}). 
The difference between the activity states is also shown on the ($q,u$) plane (Fig.~\ref{fig:QU}). 
Red triangles denote mean values, averaged over periods of the same BH state. 
There is also evidence of night-to-night variability of the  polarization during the active phase. 
 
Because of the large distance, the location close to the galactic plane, and therefore large extinction, $A_\mita{V}$ = 2.8 -- 4.4 \citep{Shahbaz2003}, one can expect a strong interstellar component in the observed polarization of \VCYG. 
In the case of \VCYG, a significant dip in polarization in the $V$ band, which appears both in the outburst and quiescence ($p_\mita{B} > p_\mita{V}$ and $p_\mita{V} < p_\mita{R}$, see Fig.~\ref{fig:V404PL}), is unusual and cannot be explained  by light scattering in the disc or synchrotron emission from a jet. 
Such a behaviour is neither consistent with dominant contribution of a single ISM component approximated by Serkowski's law \citep{Coyne1974}. 
The wavelength-dependence of the position angle of both \VCYG\ in quiescence and the companion is also not consistent with being produced in single ISM cloud.
This strongly suggests that the ISM on the line of sight has a complex structure, consisting of several screens with different particle size distributions and directions of the magnetic field. 
Their combined contribution to polarization can produce the observed wavelength dependence of $p$ and $\theta$. 
			
\subsection{Polarization of the field stars}
\label{subsec:field}
			
\begin{figure}    \includegraphics[keepaspectratio, width = 1\linewidth]{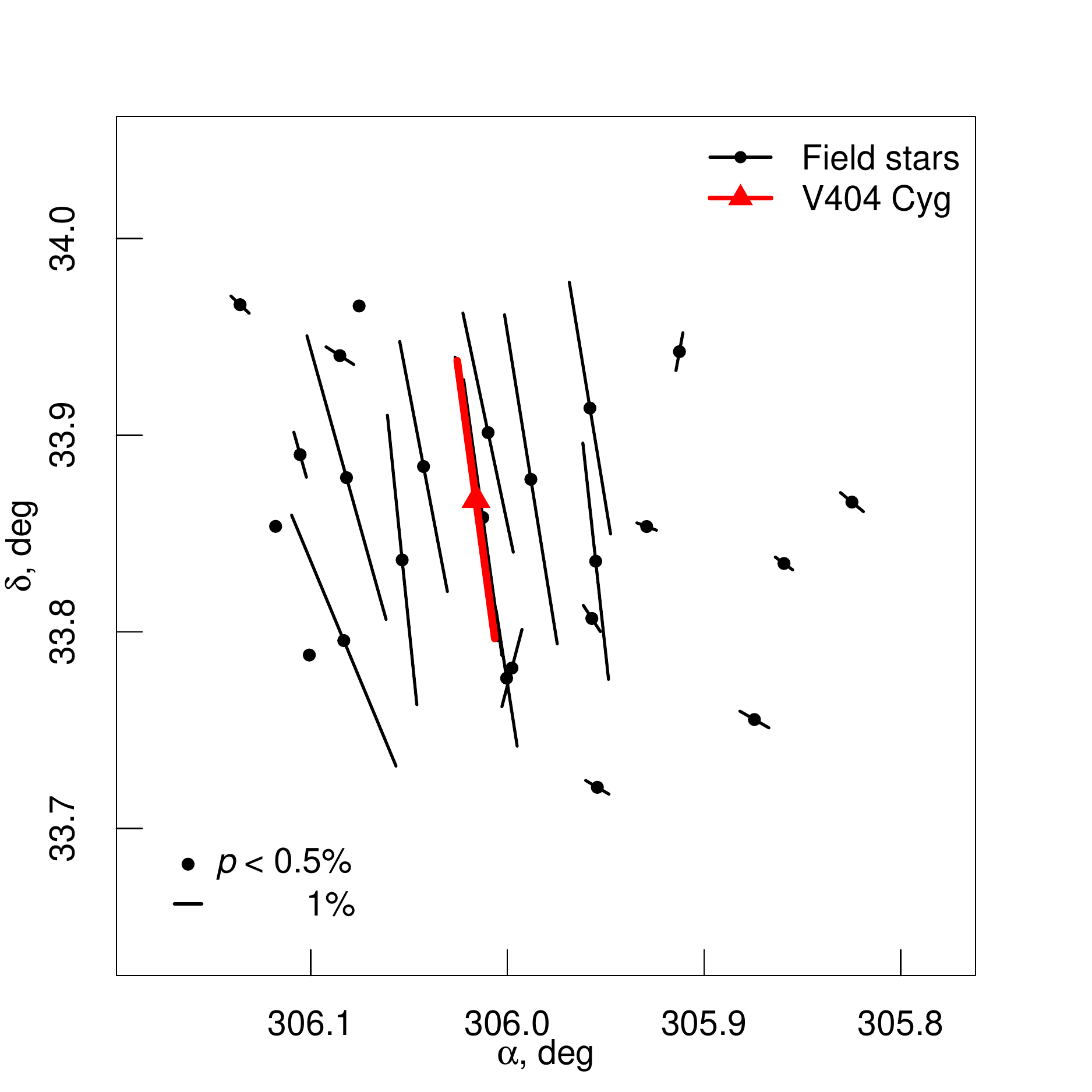}
	\caption{Polarization of the field stars (black lines, circles) as well as \VCYG\ (thick red line, triangle) in the $R$ band. North is up and east is to the left. }
	\label{fig:field}
\end{figure}

\begin{figure}	\includegraphics[keepaspectratio, width = 1\linewidth]{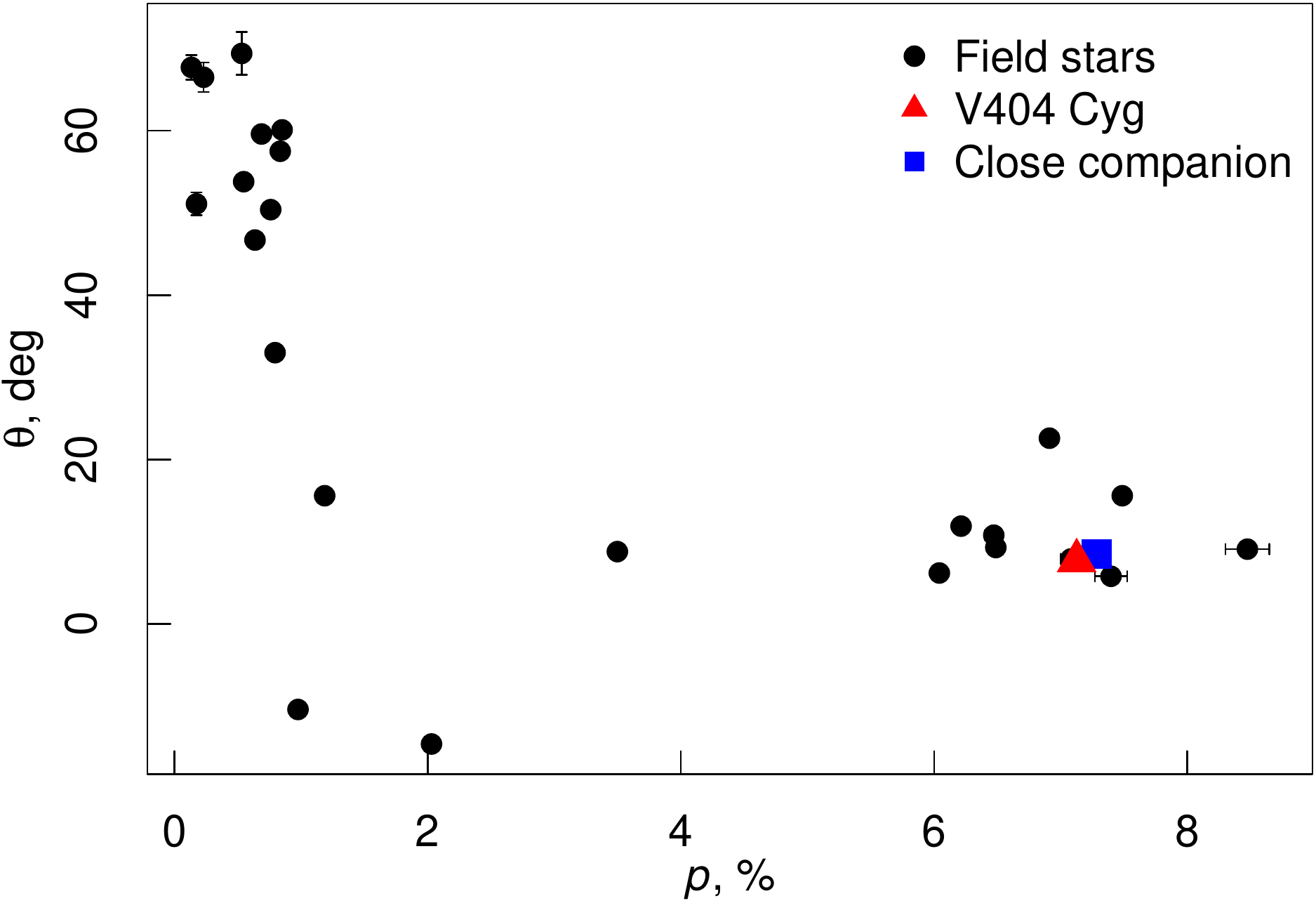}
  	\caption{Position angle -- polarization degree diagram of \VCYG\ (red triangle), close companion (blue square) and nearby stars (black circles) as observed in the $R$ band. }
  	\label{fig:PDPA}
\end{figure}

\begin{table*}
\centering
\caption{Polarimetric observations of field stars in the $B$, $V$, and $R$ bands with UH88. Errors are 1$\sigma$.}
				\label{tbl:fieldpol}
\begin{minipage}{130mm}
\begin{tabular}{ccccccc}
	\hline
    	\hline
   &  \multicolumn{2}{c}{$B$} &  \multicolumn{2}{c}{$V$} &  \multicolumn{2}{c}{$R$}  \\
	Star        &     $p$  &     $\theta$  &    $p$    &  $\theta$ &        $p$ &     $\theta$  \\
	           &   (per cent) &   (deg) &  (per cent) &  (deg) &  (per cent) &  (deg)  \\
	\hline
	   4040 &         -- &     -- &   $6.64 \pm   0.22$ &   $11.9 \pm  0.9$ &  $ 7.28\pm 0.09$ &    $8.5\pm 0.4$ \\ 
	   4041 &          -- &     -- &   $6.33 \pm 0.14$ &   $10.9 \pm 0.6$ &   $7.08\pm 0.08$ &    $7.9 \pm 0.3$ \\ 
	   4042 &        -- &     -- &   $7.09\pm    0.42$ &   $11.0 \pm 1.8$ &   $8.47\pm    0.17$ &    $9.1 \pm 0.6$ \\ 
	   4043 &   $6.92\pm 0.29$ &    $3.1 \pm     1.2$ &   $5.20 \pm 0.15$ &   $10.5\pm     0.8$ &   $6.47\pm 0.07$ &   $10.8\pm 0.3$ \\ 
	   4044 &   $5.05 \pm 0.42$ &  $178.9 \pm     2.4$ &   $5.14 \pm 0.02$ &   $10.9 \pm     0.1$ &   $6.04 \pm 0.01$ &    $6.2 \pm 0.1$ \\ 
	   4045 &   $0.77 \pm 0.02$ &   $32.2\pm     0.7$ &   $0.85 \pm 0.03$ &   $34.5\pm    1.0$ &   $0.80\pm  0.02$ &   $33.0\pm  0.9$ \\ 
	   4046 &       -- &     -- &   $5.95\pm    0.45$ &   $10.5 \pm 2.3$ &   $7.40\pm    0.13$ &    $5.8 \pm 0.5$ \\ 
	   4047 &   $6.68 \pm 0.55$ &   $18.3 \pm     2.3$ &   $6.22\pm  0.12$ &   $24.6 \pm    0.6$ &   $6.91 \pm  0.07$ &   $22.6 \pm  0.3$ \\ 
	   4048 &   $6.53 \pm 0.20$ &    $6.2 \pm     0.9$ &  $5.98\pm 0.04$ &   $11.5 \pm 0.2$ &   $6.49 \pm 0.02$ &    $9.3\pm  0.1$ \\ 
	   4049 &        -- &     -- &   $6.74 \pm   0.16$ &   $19.2\pm 0.7$ &   $7.49 \pm   0.05$ &   $15.6 \pm  0.2$ \\ 
	  40405 &   $5.85 \pm  0.36$ &    $5.8\pm     1.8$ &   $5.68 \pm 0.05$ &   $13.2   \pm   0.3$ &   $6.21\pm 0.02$ &   $11.9 \pm  0.1$ \\ 
	  40408 &   $0.41  \pm 0.07$ &   $57.4  \pm    4.9$ &   $0.36 \pm 0.03$ &   $69.8\pm     2.1$ &   $0.53 \pm 0.05$ &   $69.4\pm    2.6$ \\ 
	  40409 &   $1.22 \pm 0.02$ &   $17.1\pm    0.6$ &   $1.13\pm  0.01$ &   $16.9 \pm    0.2$ &   $1.19 \pm 0.01$ &   $15.6 \pm  0.2$ \\ 
	  40410 &   $0.18 \pm  0.01$ &   $57.1 \pm    1.9$ &   $0.17 \pm  0.01$ &   $52.3\pm     2.5$ &   $0.17  \pm   0.01$ &   $51.1 \pm     1.4$ \\ 
	  40411 &   $2.01 \pm    0.03$ &  $166.2 \pm     0.4$ &   $2.06\pm 0.03$ &  $166.7\pm     0.4$ &   $2.03 \pm  0.02$ &  $165.4\pm    0.2$ \\ 
	  40412 &   $6.28 \pm 0.12$ &    $2.5\pm    0.6$ &   $3.06 \pm  0.01$  &   $13.4 \pm    0.1$  &   $3.50 \pm   0.01$ &    $8.8 \pm  0.1$ \\ 
	  40413 &   $0.82 \pm  0.01$ &   $56.8 \pm    0.2$ &   $0.86 \pm  0.01$ &   $56.9\pm     0.5$ &   $0.84\pm  0.01$ &   $57.5 \pm    0.4$ \\ 
	  40416 &   $0.23 \pm   0.02$ &   $67.0 \pm     2.8$ &   $0.20 \pm  0.02$ &   $65.0 \pm     2.5$ &   $0.23 \pm  0.01$ &   $66.5 \pm     1.8$ \\ 
	  40417 &   $0.13 \pm  0.02$ &   $80.5 \pm    3.7$ &   $0.16 \pm  0.01$ &   $80.9 \pm     2.1$ &   $0.13 \pm  0.01$ &   $67.7 \pm     1.5$ \\ 
	  40418 &   $0.98\pm  0.02$ &  $167.7 \pm    0.7$ &   $1.01 \pm  0.02$ &  $168.8 \pm     0.5$ &   $0.98 \pm 0.01$ &  $169.6\pm  0.4$ \\ 
	  40420 &   $0.60 \pm  0.02$ &   $54.3 \pm     1.1$ &   $0.57 \pm  0.02$ &   $56.3 \pm     1.1$ &   $0.55 \pm  0.01$ &   $53.8 \pm     0.8$ \\ 
	  40421 &   $0.68 \pm  0.02$ &   $46.0 \pm    1.0$ &   $0.61 \pm  0.01$ &   $48.5 \pm     0.5$ &   $0.64 \pm  0.01$  &   $46.7 \pm  0.2$ \\ 
	  40423 &   $0.63 \pm 0.02$ &   $58.2 \pm    0.7$ &   $0.67 \pm 0.01$ &   $59.6 \pm    0.5$ &   $0.69 \pm  0.01$ &   $59.6 \pm  0.5$ \\ 
	  40425 &   $0.75  \pm  0.03$ &   $51.9\pm  1.0$ &   $0.76 \pm   0.01$ &   $51.4 \pm  0.5$ &  $0.76 \pm 0.02$ &   $50.4 \pm  0.7$ \\ 
	  40427 &   $0.82 \pm  0.03$ &   $60.0 \pm    0.9$ &   $0.85 \pm  0.01$ &   $59.5\pm     0.3$ &   $0.85 \pm  0.01$ &   $60.1\pm     0.3$ \\ 
	\hline
\end{tabular}
\end{minipage}
\end{table*}

\begin{figure}
  	\begin{minipage}{1\linewidth}
    	\includegraphics[keepaspectratio, width = 1\linewidth]{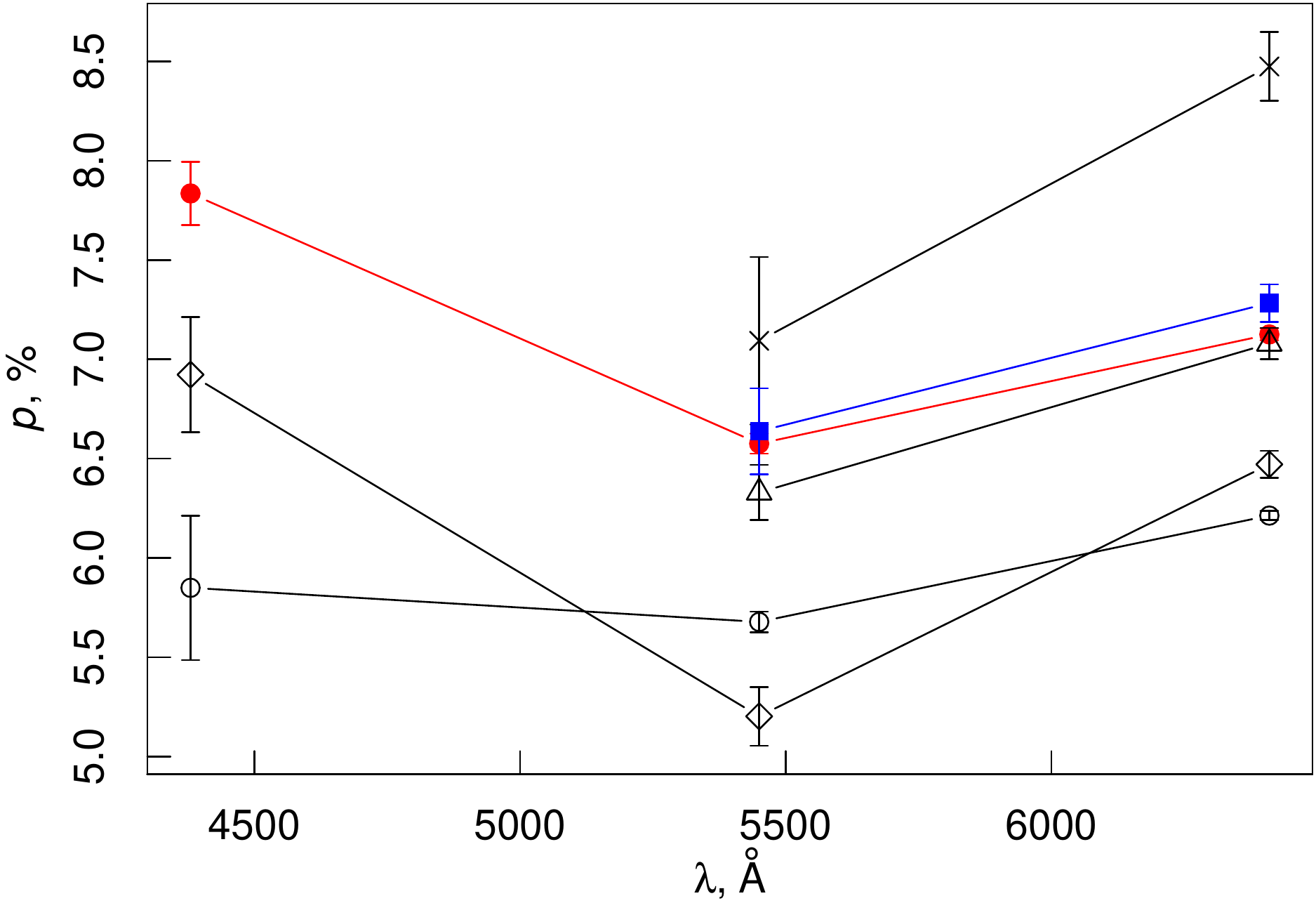}
  	\end{minipage}
	\vfill
	\begin{minipage}{1\linewidth}
    	 \includegraphics[keepaspectratio, width = 1\linewidth]{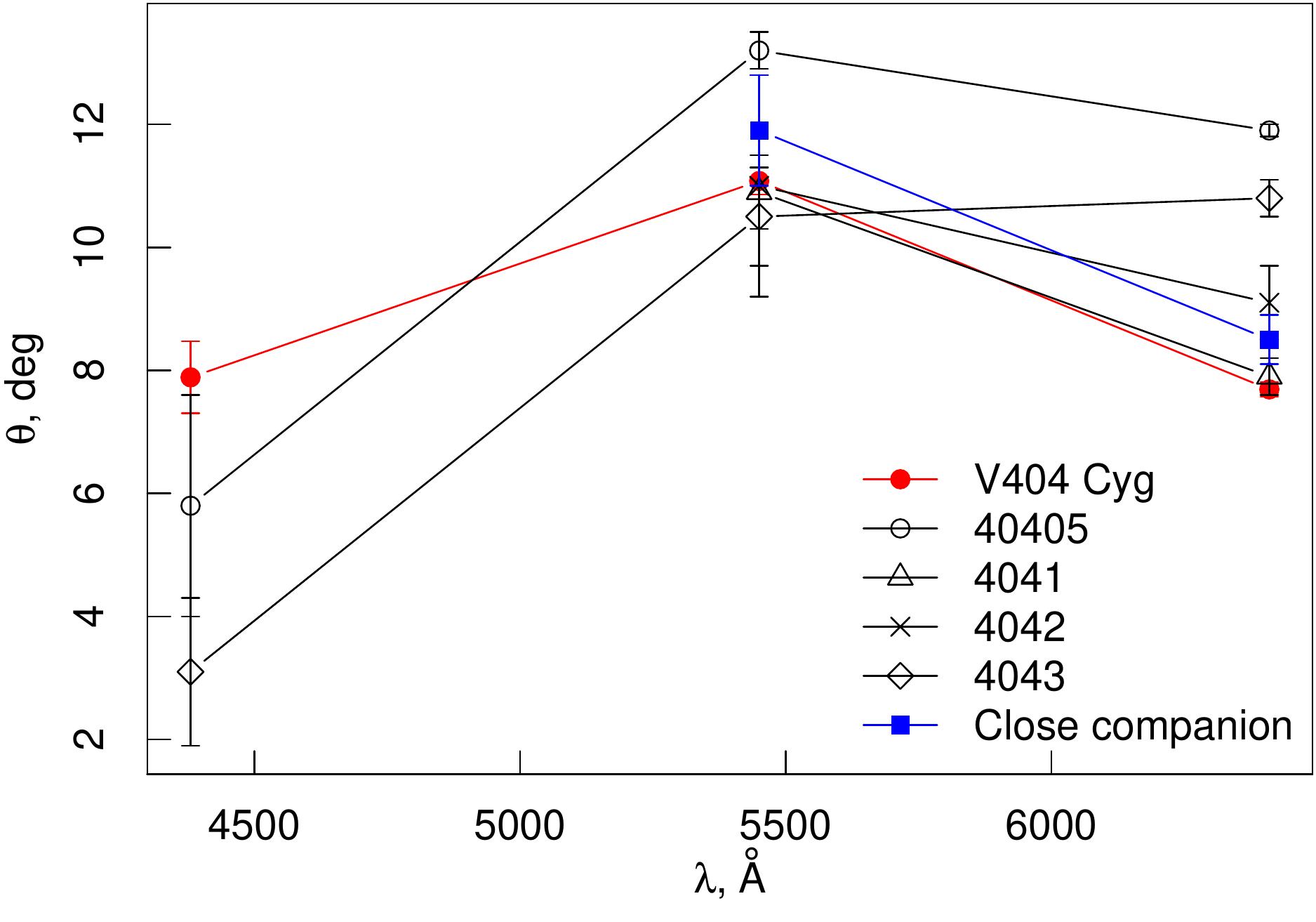}
	\end{minipage}
	\caption{Wavelength dependence of $p$ and $\theta$ of stars within 2 arcmin of \VCYG. 
Field star labels are explained in Table~\ref{tbl:fieldcoor}.}
	\label{fig:PLT2}
\end{figure}

In order to estimate the interstellar polarization in the direction of \VCYG, we have carried out a polarization study of the stars in the field of \VCYG. 
The same method has been used by \citet{Tanaka2016} and \citet{Itoh16}, but both these studies have been done mostly in the near-IR wavelength range (from $R$ to $K_{s}$). 
Our polarimeter has an important advantage of providing the data in the $B$ (for most of the observed field stars) and $V$ bands (for all stars). 
This allows us to study not only the degree and direction of the ISM polarization, but also its wavelength dependence in the ONIR range. 
Fig.~\ref{fig:field} shows polarization of stars in a region around \VCYG\ in the $R$ band. 
Most of the stars with relatively high polarization have the same $\theta$ as \VCYG\ and its close companion. 
This correlation is also seen in the $B$ and $V$ bands. 
Stars with relatively small polarization ($p \le 1$ per cent), however, show directions of polarization, which differ from $\theta$ of \VCYG\ by  $40\degr-50\degr$. 
This can be seen also in Fig.~\ref{fig:PDPA}, where $\theta$--$p$ diagram is presented. 
The same effect is clearly seen in fig.~3 of \citet{Tanaka2016}.
 
From our analysis of the polarization of the field stars we can draw several important conclusions on the nature of the ISM polarization in the direction of \VCYG:
\begin{enumerate}
\item All field stars observed by us with $p$ exceeding 5 per cent have $\theta$ in the range of $0\degr-25\degr$.	
The similar picture is seen on Figure 3 of \citep{Tanaka2016} for all stars with $p > 2$ per cent. 
The stars from our sample with $p > 5$ per cent (including \VCYG\ itself) also share peculiar wavelength dependence of polarization. 
To emphasize this, we have plotted a wavelength dependence of polarization degree and polarization angle for the stars in a small region of $2\farcm 1\times 2\farcm 1$  with the center at \VCYG\ position. 
The results are shown in Fig.~\ref{fig:PLT2}. 
\item The peculiar wavelength dependence of $p$ and $\theta$ is likely a result of the presence on the line of sight towards \VCYG\ of (at least) two screens (clouds) with different properties of the interstellar dust and different directions of the magnetic field.
Large interstellar polarization of the distant stars and \VCYG\ is produced by the propagation of light through these multiple clouds.
\item Due to the complex nature of the screens, ISM polarization towards \VCYG\ cannot be approximated by Serkowski's law. 
Therefore, the parameters of the Serkowski's law, $p_{\max}$ and $\lambda_{\max}$, derived by \citet{Shahbaz2016} from the polarization measurements of \VCYG\ in the quiescence and \citet{Itoh16} in the outburst (both without knowing the value of polarization in the blue wavelengths) are meaningless. 
\item The polarization of the close visual companion of \VCYG\ (star 4040) measured in the $V$ and $R$ bands agrees within the errors with the polarization obtained for \VCYG\ in the quiescence, which indicates that the observed polarization of \VCYG\ after the outburst is indeed predominantly interstellar. 
This fact also allows us to suggest that visual companion is spatially close to \VCYG\ and might be physically bound (with separation of $\sim$3500~AU). 
\item The results obtained by us on the ISM polarization in the direction of \VCYG\ emphasize the importance of multiwavelength polarimetry, which, preferably, should be done also in the blue passband.
\end{enumerate}

Among the field stars we have detected an object with strongly peculiar polarization, peaking in the blue. 
This star, 40412 (2MASS J20240008+3346353), is also very red, with $B-R = 4.1$~mag. 
We conclude this object is most likely intrinsically polarized. 
The sharp increase of polarization in the $B$-band is probably due to the Rayleigh scattering by dust particles in a circumstellar envelope.

\subsection{Intrinsic polarization of \VCYG}
\label{subsec:ipvcyg}

\begin{table*}
\caption{Intrinsic polarization degree and angle of \VCYG~in the $BVR$ bands for five nights during the outburst and for the combined data. \label{tbl:intr}}
\begin{minipage}{130mm}

\begin{tabular}{ccccccc}
	\hline\hline
	                      &  \multicolumn{2}{c}{$B$} &  \multicolumn{2}{c}{$V$} &  \multicolumn{2}{c}{$R$}\\
	                  MJD &              $p$ &         $\theta$ &              $p$ &         $\theta$ &              $p$ &         $\theta$\\
	                      &       (per cent) &            (deg) &       (per cent) &            (deg) &       (per cent) &            (deg)\\
	\hline
	             57195.16 &               -- &               -- &  $  1.04 \pm   0.19$ &  $ -16.7 \pm    5.2$ &  $  0.62 \pm   0.07$ &  $ -30.8 \pm    3.3$ \\ 
	             57196.15 &               -- &               -- &               -- &               -- &  $  0.78 \pm   0.23$ &  $ -22.1 \pm    8.4$ \\ 
	             57197.15 &  $  0.68 \pm   0.26$ &  $  -9.6 \pm   11.0$ &  $  1.23 \pm   0.09$ &  $  -5.4 \pm    2.1$ &  $  0.57 \pm   0.04$ &  $  -3.1 \pm    2.3$ \\ 
	             57199.14 &  $  1.73 \pm   0.57$ &  $   8.4 \pm    9.4$ &  $  0.86 \pm   0.11$ &  $  -7.1 \pm    3.7$ &  $  0.19 \pm   0.07$ &  $  -0.3 \pm    9.9$ \\ 
	             57200.14 &               -- &               -- &               -- &               -- &  $  0.25 \pm   0.10$ &  $\ \ \ 1.7 \pm   11.3$ \\ 
	      57195 --  57200 &  $  0.79 \pm   0.26$ &  $  -5.1 \pm    9.3$ &  $  1.07 \pm   0.08$ &  $  -7.0 \pm    2.1$ &  $  0.46 \pm   0.04$ &  $  -7.6 \pm    2.7$ \\ 
	\hline
\end{tabular}

\end{minipage}
\end{table*}

Polarization of \VCYG\ changes with time as the luminosity drops down after the peak around \MJD{57200} (see Figs~\ref{fig:lc} and \ref{fig:xray}). 
We split our observations into two groups, one containing observations before \MJD{57200.5} and another after that date.  
Fig.~\ref{fig:QU} show changes in the ($q,u$) parameters of \VCYG\ during the transition from one state to another. 
We have shown above that the polarization in the quiescence aligns well with the polarization of the close companion and the nearby stars (Fig.~\ref{fig:PLT2}). 
Accordingly, there seems to be an intrinsic source of polarization that is only present during an active phase, contributing about 1 per cent to the net observed polarization of \VCYG. 
Assuming that the polarization in the quiescence is produced by the ISM, we then can obtain the intrinsic polarization of \VCYG\ in the active phase by subtracting the average ($q,u$) of the quiescent state from the ($q,u$) obtained during the outburst (see Fig.~\ref{fig:QU}, bottom right panel).
				 
The error estimate of the intrinsic polarization was computed as $\epsilon_{\rm in} = \sqrt{\epsilon_{\rm out}^2 + \epsilon_{\rm qui}^2}$, where $\epsilon_{\rm out}$ and $\epsilon_{\rm qui}$ are the errors of the outburst and quiescence polarizations, respectively, and is based on a standard error propagation method. 
This formula is applied to $q$ and $u$ parameters as well as to the polarization degree $p$. 
The error on polarization angle is given by equation (\ref{eq:errortheta}).  
The average intrinsic polarization $p$ has the maximum of about $1$ per cent in the $V$ band. 
The position angles obtained in the $BVR$ bands are consistent with each other.
       
To prove the significance of change in polarization between outburst and quiescence, a statistical test should be applied. 
However, as polarization is a derived quantity and is not averaged directly, it is incorrect to test the change in polarization degree. Instead, one needs to consider averaged $q$ and $u$ Stokes parameters of combined data. 
The multivariate Hotelling's $T^2$ test \citep{Hotelling1931} can be used to estimate significance of difference between averaged $(q, u)$ parameters of two datasets. 
The smallest value of $t^2$ is expectedly obtained in $B$ band ($t^2_\mita{B} = 8$), while $t^2$ in other bands are much larger ($t^2_\mita{V} = 178$, $t^2_\mita{R} = 115$). 
The random variable  $f = \frac{n_1+n_2 - k -1}{k(n_1+n_2 - 2)}t^2$ has an $F$-distribution  with parameters $k$ and $n=n_1+n_2 - 1-k$.   
Here $k=2$ is the number of variables and $n_1, n_2$ are sizes of the combined data sets of observations made during the outburst and in quiescence, with $n_1+n_2$ being equal to $110, 122, 150$ in $B$, $V$ and $R$ bands, respectively.
The obtained $f$-values in each band are $f_B = 3.9$, $f_V = 88$ and $f_R = 57$.
The significance can  then be easily estimated using the regularized incomplete beta-function as $1 - I_x(k/2 = 1, n/2) = (1-x)^{n/2}$, where $x=kf/(n+kf)$.
Thus we get the probabilities of the outburst polarization being equal to the quiescence polarization of $0.022$, $4\times 10^{-24}$ and $5\times 10^{-19}$. 

Fig.~\ref{fig:intrT} depicts changes in the intrinsic polarization with time during the active phase  (see also Table \ref{tbl:intr}). 
The position angle $\theta$ in the $R$ band gradually changes with time from $-31\degr$ on \MJD{57195} to $\approx 0\degr$ on \MJD{57197} and further on. 
Similar variations are seen in the $V$ band, with $\theta$ varying from $-17\degr$ to $-7\degr$ 
during the same time span.
Due to the large errors on position angles it is not obvious whether the observed trends are significant. 
We tested the deviation of intrinsic angle from the weighted mean over 5 nights and obtained $\chi^2$ for $V$ and $R$ bands.  
With weighted means equal to $-7\degr [173\degr]$ and $-12\degr  [168 \degr]$ in $V$- and $R$-bands, respectively, $\chi^2_\mita{V} = 4$ with 2 degrees of freedom and $\chi^2_\mita{R} = 51$ with 4 degrees of freedom. 
These values of $\chi^2$ correspond to the probabilities that $\theta$ does not change equal to 0.135 and $2\times 10^{-10}$ in $V$- and $R$-bands, respectively. 
Thus variations in $V$ are not significant,  but changes in $R$ are highly significant. 

\begin{figure}
		 	\begin{minipage}{1\linewidth}           
					 \includegraphics[width = 1\linewidth, keepaspectratio]{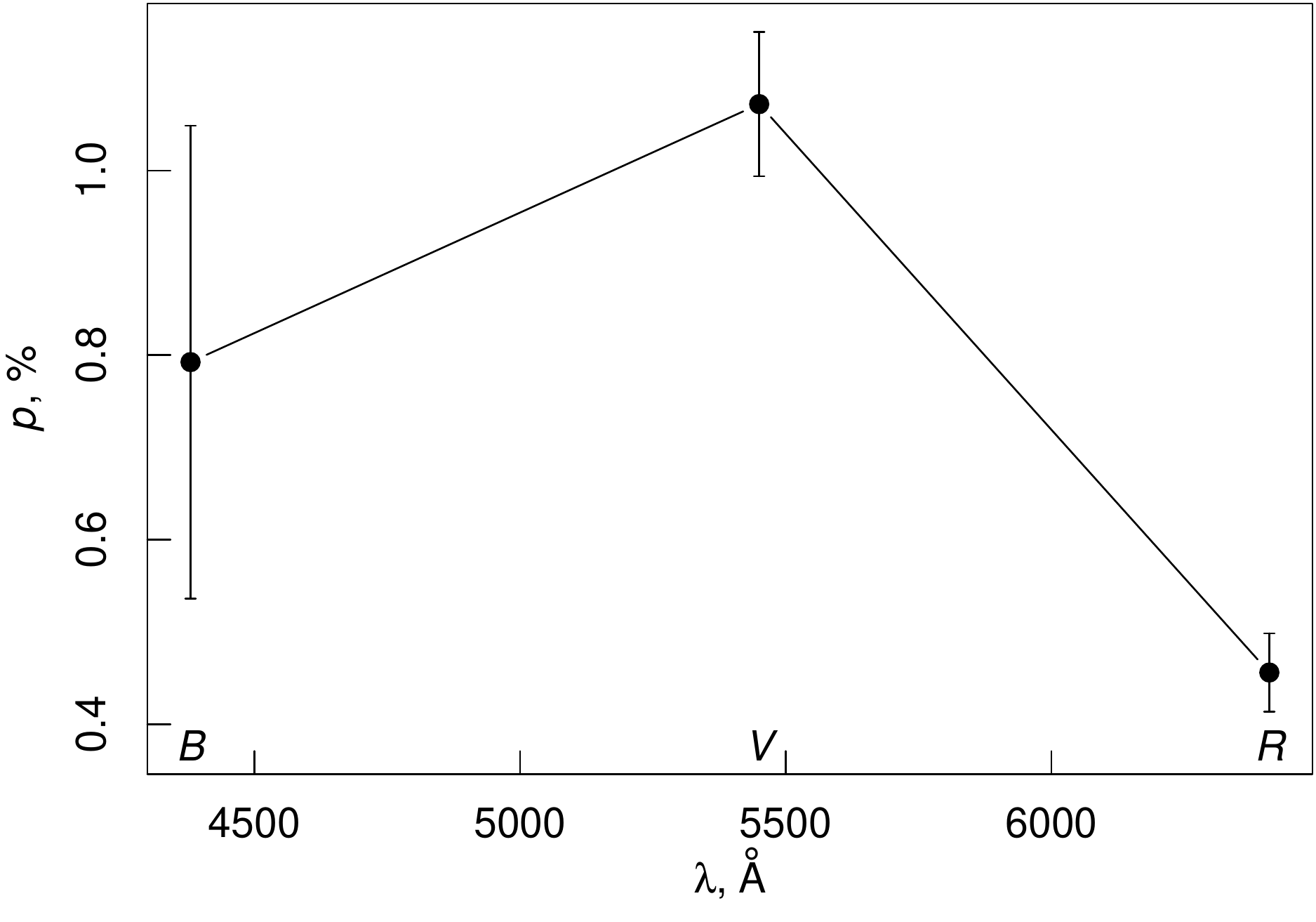}
				\end{minipage}
				\vfill
			 	\begin{minipage}{1\linewidth}
			 		\includegraphics[width = 1\linewidth, keepaspectratio]{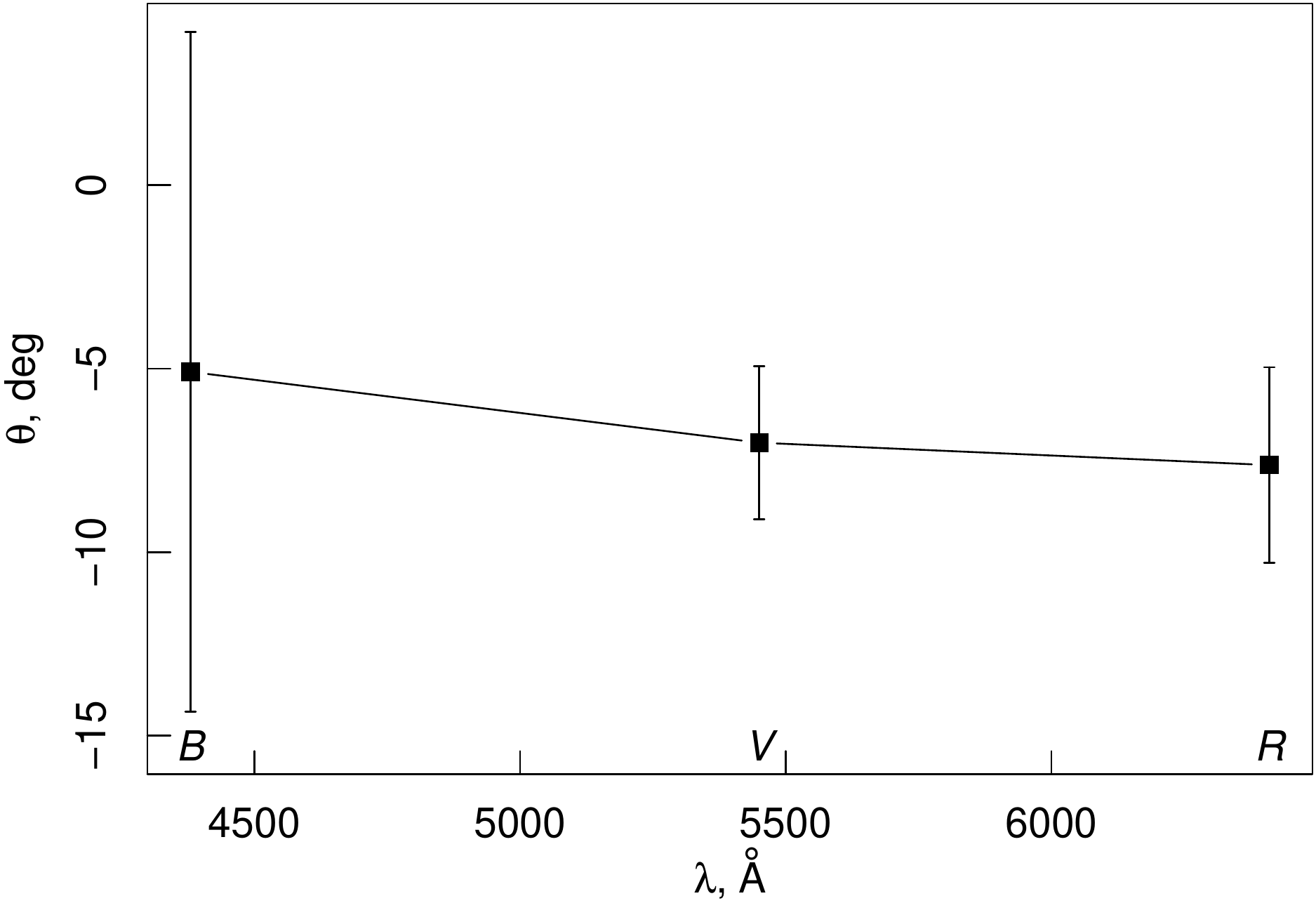}
			 	\end{minipage}
				 \caption{Intrinsic average polarization degree $p$ (top panel) and polarization position angle $\theta$ (bottom panel) of \VCYG\ as functions of the wavelength. }
				 \label{fig:intr}
			 \end{figure}

			 \begin{figure}
 			 	\begin{minipage}{1\linewidth}		
     \includegraphics[keepaspectratio, width = 1\linewidth]{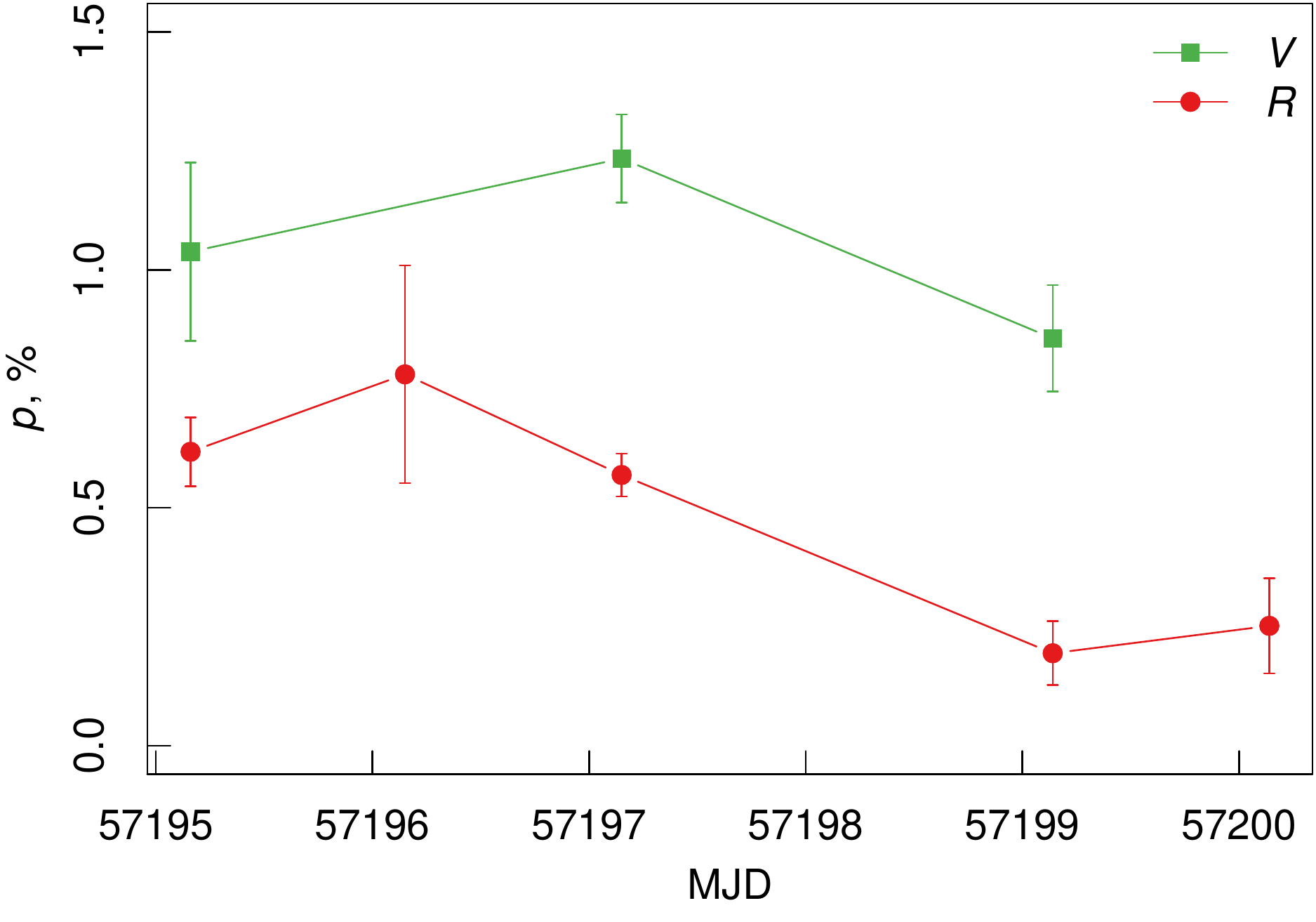}
	 			 \end{minipage}
	 			 \begin{minipage}{1\linewidth}
                    \includegraphics[keepaspectratio, width = 1\linewidth]{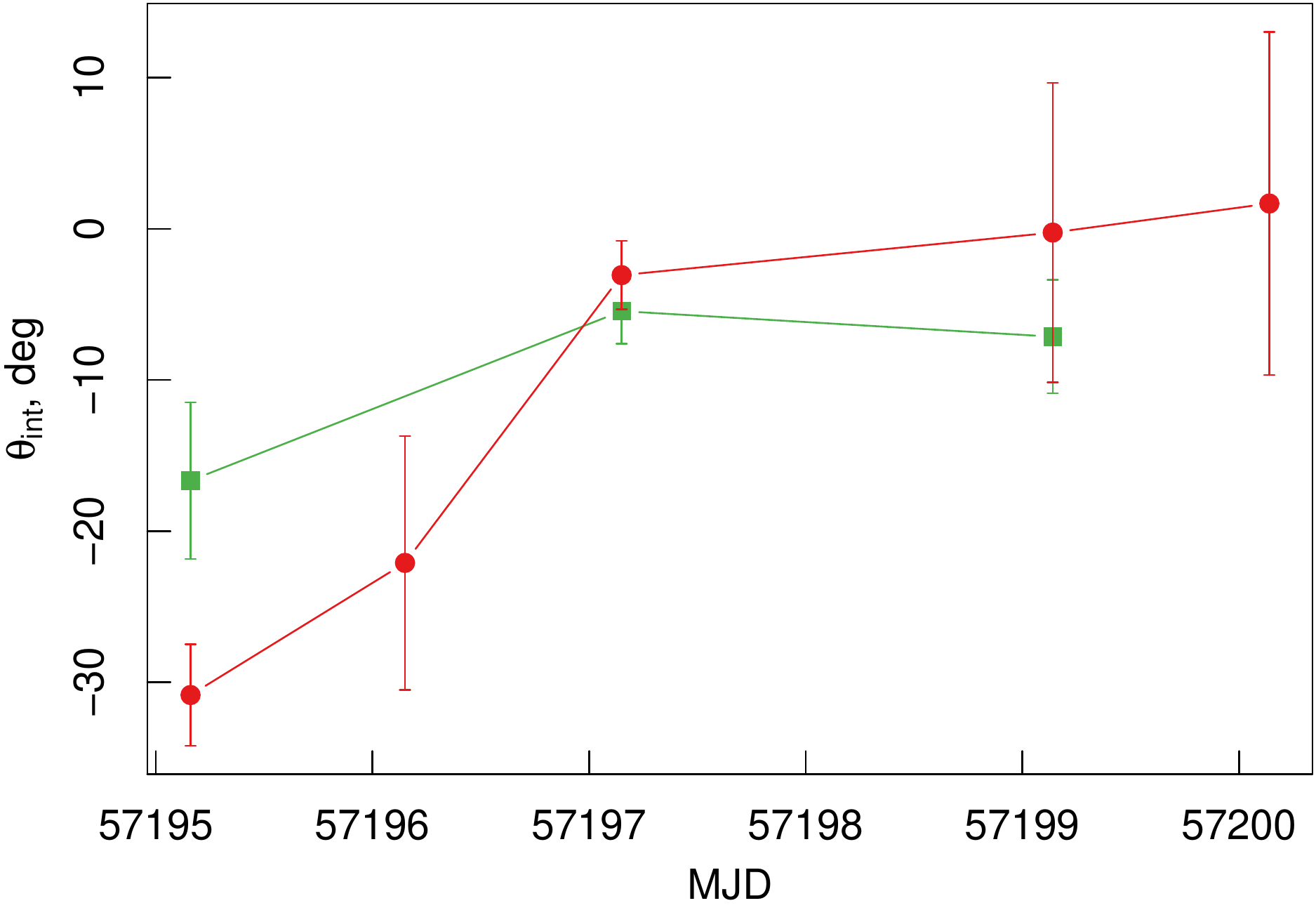}
	 			 \end{minipage}
	 			 \caption{Intrinsic polarization $p$ (top panel) and  position angle $\theta$ (bottom panel) as functions of time.}
 			 \label{fig:intrT}
			 \end{figure}

\subsection{Summary of the main results}
                
Our ONIR polarization observations of \VCYG\ during the June 2015 outburst and the quiescence have showed that there is a highly significant difference in the Stokes parameters ($q,u$) between the activity states in all three passbands used (Fig.~\ref{fig:QU}, lower right panel). 
The variable polarization of \VCYG\  indicates there is an intrinsic source of polarization responsible for the observed changes in $p$ and $\theta$. 
As was discussed in Section~\ref{subsec:field}, the polarization profile of the field stars in the direction of \VCYG\ and the visually close (1\farcs4) companion match the profiles of \VCYG\ during its quiescence, indicating that the quiescence polarization is of interstellar origin.                

The intrinsic source of polarization, which was present during the active phase, contributes 0.5--1.0 per cent to the overall observed polarization of the object, reaching the maximum in the $V$ band. 
The polarization angle of the average intrinsic component does not significantly depend on wavelength. 
However, there is some evidence of variation of the intrinsic polarization during the active period. 
Although the degree of the intrinsic polarization, $p$, seems to remain almost constant in both $V$ and $R$ bands, the position angle $\theta$ shows gradual changes with time from
$-17\degr [163\degr] \pm 5\degr$
to 
$-5\degr [175 \degr] \pm 2\degr$
in $V$
and very significant variations from
$-31\degr [149\degr] \pm 3\degr$
to
$-3\degr [177\degr] \pm 2\degr$
in $R$ 
in two days (\MJD{57195--57197}) with $\theta_\mita{R}$ being nearly constant after that until \MJD{57200} (see Table \ref{tbl:intr}).  
			
\section{Discussion}
\label{sec:disc}

\subsection{Comparison with other observations}

Several reports of polarization observations of \VCYG\ during its active phase have been published. 
However, the time intervals when the observations were made differ, and it is not always possible to directly compare the results. 
Furthermore, because \VCYG\ appears quite red, accurate data are reported mostly in the $R$ band. 
                
\citet{Tanaka2016} observed \VCYG\ starting from \MJD{57190} for 4 nights using two instruments, in the $R$ and the $K_\mrm{s}$ bands. 
On  night of \MJD{57194}, observations in the $VR_{\mrm{c}}I_{\mrm{c}}JHK_{\mrm{s}}$ have been made. 
The data obtained show that there was no significant variability in $p$ and $\theta$ in the $R$ band, at least during the nights of \MJD{57193} and \MJD{57194}. 
Polarization degree and polarization angle obtained at \textit{Pirka}/MSI
($ p_\mita{R} = 7.77 \pm 0.01$ per cent, $\theta_\mita{R} = 6\fdg 19 \pm 0\fdg 03$  on \MJD{57193}) 
are consistent with our results (Fig.~\ref{fig:lc}ab). 
Polarization in the $V$-band of $8.9 \pm 0.1$ per cent, however, measured on \MJD{57194} \citep{Itoh16} is significantly higher than
$p_\mita{V} = 7.22 \pm 0.18$ per cent
measured by us on the next day.
An extensive polarization survey of the field stars in the near-IR, with the few stars measured also in the $V$-band, have been done by \citet{Itoh16}. 
Main results of this survey agree with our data obtained on the field stars very well (compare fig.~3 from \citealt{Tanaka2016} with our Fig.~\ref{fig:PDPA}). 
The authors concluded that \VCYG\ polarization is predominantly of interstellar origin, with possible contribution from intrinsic component of not more than few percent, which again agrees with our estimations of intrinsic polarization, $p_\mita{R} = 0.46\pm 0.04$ to $p_\mita{V} = 1.07\pm 0.08$ per cent. 
                
\citet{Shahbaz2016} observed variable polarization during the active phase. 
However, even though their observations at the TNG were made on the same night \MJD{57197} as one of our sets, we are unable to compare them directly with our measurements, because the original polarization data obtained in the $r^\prime$ band were not given in their paper. 
The intrinsic polarization derived by \citet{Shahbaz2016} is 3.5--4.5 per cent, which is significantly higher than the intrinsic $R$ band polarization 
($0.46\pm 0.04$ per cent) obtained by us. 
If one takes into account the value of the subtracted ISM polarization adopted by \citet{Shahbaz2016},
$p_\mita{R}^\mrm{IS} = 7.41\pm 0.32$ per cent
and
$\theta_\mita{R}^\mrm{IS} = 7\fdg 2 \pm 1 \fdg 1$
(see their Table 1), and add  the respective Stokes parameters to the ones obtained from polarization data (shown on their fig.~3),
the estimate of the observed (measured at the TNG) polarization  is $p_\mita{R}^\mrm{obs} = 10-12$ per cent.
 
This is much higher than polarization in the R-band measured by us ($p_\mita{R} = 7.66 \pm 0.03$ per cent on \MJD{57197}). 
We cannot explain this difference between our and TNG measurements made on the same night \MJD{57197}. 
We can only note that TNG polarimeter, installed in the Nasmyth focus, has the variable instrumental polarization of about 2--3 per cent which must be carefully calibrated and removed \citep{giro03}. 
Our instrument, the Dipol-2, benefits from being free of any instrumental polarization at the level of less than 0.01 per cent.
 
The observed polarization of \VCYG, measured at the TNG, also appears to be much higher than all other measurements obtained in the outburst by other observers: $p_{\mita{R}} = 7-8$ per cent (see Section 3.1 in \citealt{Shahbaz2016} where these values of polarization are given). 
At the same time, the intrinsic mean polarization angle  $\theta= -9\degr [171\degr]$ presented in \citet{Shahbaz2016}  is consistent with what we observed: from 
$-5\degr [175\degr] \pm 9 \degr$ in $B$
to $-8\degr [172\degr] \pm 3 \degr$ in $R$ (see Fig.~\ref{fig:intr}).

\citet{Lipunov2016} have reported a detection of variable polarization in \VCYG\ with the amplitude of 4--6 per cent over a timescale of approximately 1 hr on two epochs during the outburst. 
The setup which was used by for polarimetry of \VCYG\ with the MASTER robotic net does not allow to measure a true polarization (i.e. Stokes parameters $q$ and $u$) with a single telescope, but rather a difference in fluxes recorded with two different CCD detectors on two different telescopes, equipped with orthogonally oriented polarizing filter.
This difference, as they argue, allows to measure a lower limit of the true polarization. 
Obviously, this setup makes the accurate calibration of instrumental polarization, which can be field dependent, literally impossible. 
It  can neither effectively eliminate the influence of sky polarization, clouds, etc. 
Therefore \citet{Lipunov2016} must rely on the comparison of polarization of \VCYG\ with the field-star polarization, assumed to be of interstellar origin and therefore constant.
We note that  in addition to  \VCYG\  there is at least one field star which clearly shows fast polarization variability with the amplitude of
at least 2 per cent, which exceeds the error bar by  a factor of $\sim$10 (see the uppermost green symbols for the field stars in their fig. 1).
We can compare this with the variablity of \VCYG\ itself, shown on the same figure: the amplitude is $\sim$4 per cent, and the error bar of $>1$ per cent (i.e. exceeding 10 times the error for the abovementioned field star). 
It is thus obvious that there is at least one more star in the field which shows very fast variability at much higher confidence level than  \VCYG. 
This raises the doubts on the quality of the data coming from MASTER polarimetry and the reality of the claimed polarization flares.

\subsection{Interpretation}
 
The data show variations of the  polarization angle $\theta$ with time which is the most pronounced in the $R$-band (see Fig.~\ref{fig:intrT}). 
Interestingly, the  range of  measured $\theta$ (from  $-30\degr$ to  $0\degr$)  coincides with the position angles of jet ejections that are resolved by the VLBI  (J. Miller-Jones et al. 2017, in prep.).  
Because our observations are rather short and not strictly simultaneous with the VLBI observations, we cannot claim here that there is a definite relation between the polarization position angle and the jet directions.  
However, the coincidence is intriguing.  

Do these observations then imply that ONIR polarized emission is produced by the jet? 
Not at all. 
The intrinsic spectral polarization profile of \VCYG\ shows a maximum in the $V$-band with $p_\mita{V}$ reaching 
$1.23\pm0.09$ per cent on \MJD{57197} and polarization drops towards $R$-band by a factor of two 
to $p_\mita{R} = 0.57\pm0.05$ per cent measured on the same day (see Fig. \ref{fig:intrT}).
If the jet were the source of polarized flux, we would expect exactly opposite behaviour with the polarization growing toward the red just because the ONIR jet spectrum is much softer (redder) than the accretion disc spectrum \citep[see e.g.][]{Gandhi11,Russell13breaks}. 
Furthermore, the radio flux observed by RATAN and NOEMA  on \MJD{57199--57200} is very low  \citep[see fig. 10 in][]{Rahoui17} implying the jet contribution of at least two orders of magnitude below the observed flux in the ONIR range.
This clearly shows that the jet cannot be responsible for observed polarization.

On the other hand, the observed wavelength dependence of polarization resembles that of Be stars, where  electron scattering in a circumstellar disc leads to the intrinsic polarization of about 1.0--1.5 per cent  \citep{Poeckert1979}.  
Although Thomson scattering opacity is independent of the wavelength, the degree of polarization is reduced and the spectral shape of the polarized flux is modified by hydrogen absorption in Balmer and Paschen series \citep{Nagirner62,Poeckert1978}. 
Also dilution by unpolarized free-free  and dust emission from the envelope around the accretion disc may contribute towards the red;   
such a component has indeed been observed as an NIR excess above the Rayleigh-Jeans tail  \citep{Rahoui17}.  

Studies of Be stars show that the degree of intrinsic polarization decreases in hydrogen lines and across hydrogen series limits. 
Our observations in the $BVR$ bands are made between the Balmer and Paschen limits, 
but H$_{\gamma}$ ($\lambda = 4341$ \AA) and H$_{\alpha}$ ($\lambda = 6563$ \AA) lines are within the $B$- and  $R_{\rm c}$-filter, respectively. 
The lines may be partially responsible for the decrease of polarization in these bands. 
However, during the outburst the line contribution to the continuum flux does not exceed 10 per cent \citep{MunozDarias16Nat}, implying that the observed form of $p(\lambda)$ is not related to the lines. 
If our interpretation of the wavelength dependence is correct,  we expect that the polarization degree  will be strongly reduced to the blue of the Balmer limit and slightly increased to the red  of the Paschen limit \citep{Poeckert1978}. 
Decrease of polarization in the $B$-band could also be partly due to the relatively low signal-to-noise ratio of our polarimetry in the blue wavelengths.  
               
Geometrical properties of the emitting region and its optical depth also play a significant role in polarizing the radiation. 
The degree of polarization depends on the ellipticity of the source. 
Spherical regions produce no polarization, while slab-like structures may produce  significantly linearly polarized radiation up to tens of per cent, depending on optical depth and the role of electron scattering. 
For optically thick, electron-scattering dominated envelopes the intrinsic polarization is $<12$ per cent and is parallel to the  disc plane (i.e. perpendicular to the disc normal) and the equatorial plane of a flattened ellipsoidal shell \citep{Cha47,Cha60,Sob49,Sob63}. 
In the optically thin case,  polarization is  parallel to the  disc normal and can reach higher values \citep{ST85,BP99,VP04}.  
Intermediate ellipsoids introduce small polarizations of the order of a few percent \citep{DGS95,Gnedin97}, which is the case of Be stars.

The ONIR emitting region, of course, does not need to have an ellipsoidal shape. 
A slow accretion disc wind may be concentrated to the equatorial plane. 
The presence of such a wind  occupying a rather large solid angle is supported by the  observed very broad P~Cyg profiles in the optical \citep{MunozDarias16Nat} as well as in the X-ray \citep{King15} lines.   
In this case,  the polar scattering region which induces polarization parallel to the disc plane might be completely missing.
This  would result in the net polarization parallel to the disc normal. 

Scattering in the polar region may also induce parallel polarization if the outflow is mildly relativistic  \citep{Beloborodov98,BP99}, because relativistic aberration  causes  a limb brightening  of  the disc radiation in the outflow frame.   
The wavelength dependence of the polarization degree can be reproduced in this case if the scattered radiation  is blue-shifted relative to the seed disc radiation (which is actually expected if the electron temperature or the bulk motion of the outflow is mildly relativistic) and there is a diluting red unpolarized component coming from the wind \citep{Rahoui17}.

The impact of absorption opacity can also be significant, reducing importance of scattering and leading to a rotation of the polarization plane by $90\degr$ (relative to the case of pure scattering) at some (typically close to the slab normal) viewing angles  known as Nagirner effect \citep{Nagirner62,DGS95,Gnedin97}. 
Thus both optically thin, electron-scattering-dominated envelopes and those with significant absorption produce polarization parallel to the disc normal (and therefore the jet direction).  
A small observed value of polarization and a strong wavelength dependence, however,  imply that it is the interplay of scattering and absorption in a flattened envelope that  might be responsible for the observed properties. 

Because of the high source luminosity likely above the Eddington at least on some occations during the outburst (see Fig.~\ref{fig:xray}a and \citealt{Segreto15,Rodriguez15b,Rodriguez15a}) and  the evidence of a strong wind \citep{King15,MunozDarias16Nat}, it is very likely  that a significant fraction of the ONIR radiation is produced  in the  outflow \citep{PLF07}. 
In many respects \VCYG\ is similar to V4641~Sgr, which also showed bright optical emission  which was  interpreted as evidence of a large envelope enshrouding the central BH accreting at super-Eddington rate \citep{Revnivtsev02a,Revnivtsev02b}.  
The strong iron line observed in both cases \citep{King15,Revnivtsev02b} implies that the X-rays from the central source have to diffuse through an envelope. 
The picture thus emerges that the best model describing the orientation of the  polarization angle along the jet direction and the wavelength dependence of the polarization degree together with the  sharp variability in the optical and X-ray  bands is the equatorial, clumpy, not fully ionized wind. 
Alternatively, the polarized signal may also be produced in the polar mildly relativistic outflow 
in combination with a diluting red unpolarized component.  
The data at hand do not allow to  differentiate between the models.

\section{Conclusion} \label{sec:concl}
        
We observed \VCYG\ polarization during the 2015 June  outburst and in the quiescence, in the broad $BVR$ bands. 
Variable source polarization was detected in the active phase. 
We carried out a survey of surrounding field stars in order to determine the properties of interstellar polarization in the vicinity of \VCYG. 

The magnitude and direction of polarization of \VCYG\ in the $V$ and $R$ bands in quiescence are very close to those of the visually close (1\farcs4) companion. 
We consider this as an evidence that these two stars are spatially close and might be physically bound. 
Polarization direction and wavelength dependence of \VCYG\  in quiescence are similar to those of the nearby field stars. 
Therefore, it must be predominantly of interstellar origin. 
The obtained intrinsic polarization during outburst shows variability with the polarization angle changing on the time scale of 2--3 days from $-31\degr$ to $\sim 0\degr$, in the range of the position angle of the multiple ejections observed with the VLBI on the same days. 
The wavelength dependence of the intrinsic polarization, peaking in the $V$ band and decreasing towards the red, indicates that the source of the intrinsic polarization most likely is not the synchrotron emission from the jet. 
On the other hand, the data are consistent with the interpretation that polarization arises in a flattened envelope (outflow) that fully enshrouds the accretion disc around the BH or,  alternatively, by scattering of the disc radiation in the mildly relativistic polar outflow. 

\section*{Acknowledgements}

This research was supported by the University of Turku Graduate School  in  Physical and Chemical Sciences (IK), the ERC Advanced grant HotMol (ERC-2011-AdG291659; AVB, VP), the Foundations' Professor Pool, the Finnish Cultural Foundation, the Academy of Finland grant 268740, and the National Science Foundation grant PHY-1125915 (JP).
This article is based on observations made in the Observatorios de Canarias del IAC with the WHT telescope operated on the island of La Palma by the ING Group, and the KVA 60 cm telescope (Kungliga Vetenskapsakademien, Sweden), in the Observatorio del Roque de los Muchachos. The authors are grateful to the Institute for Astronomy, University of Hawaii (IfA), for the allocation of IfA time for the Dipol-2 observations at the UH88 telescope. 
We thank James Miller-Jones for sharing the VLBI data before the publication, Harry Lehto for help with the statistics  and the anonymous referee for helpful comments. 




	\bibliographystyle{mnras}
	\bibliography{Cyg} 




	\bsp	
	\label{lastpage}
\end{document}